\pdfoutput = 1

\documentclass[12pt,a4paper]{article}

\usepackage{cite}
\usepackage{placeins}

\usepackage[utf8]{inputenc}

\usepackage{amsmath}
\usepackage[dvipsnames]{xcolor}
\usepackage{amssymb}
\usepackage{graphicx}
\usepackage[colorlinks=true, allcolors=Blue]{hyperref}
\usepackage{caption}
\usepackage{subcaption}
\usepackage{braket}
\usepackage{slashed}
\usepackage{float}
\usepackage{multirow}
\usepackage{cite}

\mathcode`l="8000
\begingroup
\makeatletter
\lccode`\~=`\l
\DeclareMathSymbol{\lsb@l}{\mathalpha}{letters}{`l}
\lowercase{\gdef~{\ifnum\the\mathgroup=\m@ne \ell \else \lsb@l \fi}}%
\endgroup

\begin{document}
\begin{titlepage}

\vspace*{-0.7truecm}
\begin{flushright}
Nikhef-2021-017
\end{flushright}

\vspace{1.6truecm}

\begin{center}
\boldmath
{\Large{\bf  Using \boldmath$B^0_s\to D_s^\mp K^\pm$ Decays as a Portal to \\ 

\vspace*{0.2truecm}

New Physics }
}
\unboldmath
\end{center}

\vspace{0.8truecm}

\begin{center}
{\bf Robert Fleischer\,${}^{a,b}$ and  Eleftheria Malami\,${}^{a}$}

\vspace{0.5truecm}

${}^a${\sl Nikhef, Science Park 105, NL-1098 XG Amsterdam, Netherlands}

${}^b${\sl  Department of Physics and Astronomy, Vrije Universiteit Amsterdam,\\
NL-1081 HV Amsterdam, Netherlands}

\end{center}

\vspace*{1.7cm}

\begin{abstract}
\noindent
The system of $B^0_s\to D_s^\mp K^\pm$ decays offers a theoretically clean determination of the angle $\gamma$ of the Unitarity Triangle. A corresponding LHCb analysis resulted in a surprisingly large result, which is in tension with other determinations and global analyses of the Unitarity Triangle. Paying special attention to discrete ambiguities, we confirm this picture and resolve a final ambiguity. Moreover, we extract the branching ratios of the underlying $\bar{B}^0_s\to D_s^+K^-$ and $\bar{B}^0_s\to D_s^- K^+$ modes. Combining them with information from semileptonic $B_{(s)}$ decays, we arrive at another puzzling situation, which we obtain also for other decays with similar dynamics. These patterns could be footprints of New Physics in the $b\to c \bar{u} s$ and $b \to u \bar{c} s$ processes which govern the dynamics of the $\bar{B}^0_s\to D_s^\mp K^\pm$ channels. Employing a model-independent parametrisation, we present a strategy to reveal such effects. Applying it to the present data, we obtain strongly correlated New-Physics contributions with potentially large CP-violating phases. We find that new contributions sizeably smaller than the Standard Model amplitudes could actually accommodate the current data. This strategy offers an exciting probe for new sources of CP violation at the future high-precision frontier of $B$ physics. 
\end{abstract}

\vspace*{2.1truecm}

\vfill

\noindent
September 2021

\end{titlepage}

\newpage

\thispagestyle{empty}
\vbox{}
\newpage

\setcounter{page}{1}

\section{Introduction}
The decays $\bar B^0_s\to D_s^+K^-$ and $B^0_s\to D_s^+K^-$ with their counterparts decaying into the CP-conjugate final state $D_s^-K^+$ provide a particularly interesting laboratory for the exploration of CP violation \cite{ADK,RF-BsDsK,DeBFKMST}. In the Standard Model (SM), these channels originate only from tree-diagram-like topologies caused by $b\to c \bar{u} s$ and $\bar{b} 
\to \bar{u} c \bar{s}$ quark-level processes, respectively. Thanks to $B^0_s$--$\bar B^0_s$ mixing, interference effects between the different decay paths into the $D_s^\mp K^\pm$ final states arise, thereby providing observables which allow the determination of a CP-violating phase $\phi_s+\gamma$. Here $\gamma$ is the usual angle of the Unitarity Triangle (UT), while $\phi_s$ is the CP-violating $B^0_s$--$\bar B^0_s$ mixing phase. The key feature of this method is that non-perturbative hadronic matrix elements of four-quark operators cancel in certain combinations of CP-violating observables, thereby resulting in a theoretically clean determination of $\phi_s+\gamma$ within the SM. The value of $\gamma$ can then be extracted with the help of the measured value of $\phi_s$, which can be determined through CP violation in $B^0_s\to J/\psi \phi$ decays and channels with similar dynamics \cite{DDF,Dunietz:2000cr,Faller:2008gt,DeBF,Barel:2020jvf}.

The LHCb collaboration has performed an interesting study of CP violation in the system of the 
$\bar B^0_s\to D_s^\mp K^\pm$ and $B^0_s\to D_s^\mp K^\pm$ decays, reporting the result 
\begin{equation}\label{LHCb-gam}
\gamma=\left(128^{+17}_{-22}\right)^\circ
\end{equation}
 modulo $180^\circ$ from a fit to the data, where the uncertainty contains both statistical and systematic contributions \cite{LHCb-BsDsK}.  Strategies utilising decays of the kind $B\to D K$, which also originate only from tree-level topologies in the SM, result in values of $\gamma$ in the regime of $70^\circ$ \cite{Amhis:2019ckw,PDG,LHCb:2021dcr}, 
 which is also consistent with global analyses of the UT \cite{CKMfitter,UTfit}. 
Consequently, despite the significant uncertainty, the result for $\gamma$ in Eq.~(\ref{LHCb-gam}) with its central value 
much larger than $70^\circ$  is intriguing. Could it actually indicate new sources of CP violation originating from physics beyond the SM? In order to shed light on this question, we will have an independent look at the extraction of $\gamma$, paying special attention to discrete ambiguities and their resolution \cite{RF-BsDsK,DeBFKMST}.  We obtain a picture fully consistent with Eq.~(\ref{LHCb-gam}), and resolve the final ambiguity of modulo $180^\circ$, corresponding to $\gamma=\left(-52^{+17}_{-22}\right)^\circ$.

Further puzzling patterns arise in $\bar B^0_d\to D_d^+ K^-$, $\bar B^0_d\to D_d^+ \pi^-$ and $\bar B^0_s\to D_s^+ \pi^-$ modes,  
which are tree decays with dynamics similar to $\bar B^0_s\to D_s^+K^-$. 
Their branching ratios are found experimentally to be too 
small with respect to QCD factorisation \cite{FST-BR,Bordone:2020gao}, which is expected to work very well in this decay class \cite{Beneke:2000ry}. This feature has recently been addressed within New Physics (NP) analyses \cite{Iguro:2020ndk,Cai:2021mlt,Bordone:2021cca}; NP effects in non-leptonic tree-level decays of $B$ mesons were also studied in
Refs.\ \cite{Brod:2014bfa,Lenz:2019lvd}.
 What is the corresponding situation for the branching ratios of the $\bar B^0_s\to D_s^+K^-$ and $B^0_s\to D_s^+K^-$ modes? We will extract these quantities from the experimental data. Complementing the
branching ratios with information from semileptonic $B_{(s)}$ decays, we determine parameters characterising factorisation from the data.
For the $\bar B^0_s\to D_s^+K^-$ channel, we obtain a picture similar to the $\bar B^0_d\to D_d^+ K^-$, $\bar B^0_d\to D_d^+ \pi^-$ and 
$\bar B^0_s\to D_s^+ \pi^-$ modes, showing tension with QCD factorisation as well. Interestingly, we find such a pattern -- 
although with larger uncertainties -- also for $B^0_s\to D_s^+K^-$ and $B^0_d\to D_s^+\pi^-$, having similar dynamics.

In view of these puzzles in the data for the $B^0_s\to D_s^\mp K^\pm$ decays, we allow for CP-violating NP contributions and present a new strategy to reveal such effects. The corresponding model-independent formalism allows a transparent analysis to constrain possible NP effects with potentially large CP-violating phases. We go beyond assumptions made in the LHCb analysis 
 \cite{LHCb-BsDsK}, point out SM relations and propose new measurements and aspects for future experimental studies. 

In our analysis of the $B^0_s\to D_s^\mp K^\pm$ system, we first assume the SM. The outline of this paper
is as follows: in Section~\ref{sec:CPV}, we have a closer look at CP violation and discuss the determination of $\phi_s+\gamma$ from the measured observables, paying special attention to discrete ambiguities and their resolution. In Section~\ref{sec:BR}, we extract the branching ratios of the individual $B^0_s\to D_s^+K^-$ and $B^0_s\to D_s^-K^+$ channels from the experimental data. Complementing them with information from semileptonic $B_{(s)}$ decays, we determine hadronic parameters, allowing a comparison with QCD factorisation. In view of the puzzling patterns arising in these studies, we extend the analysis to include contributions from physics beyond the SM in Section~\ref{sec:NP}. Here we present a model-independent framework to reveal possible NP effects from the data, and apply it to the currently available measurements. Finally, we summarise our findings and conclusions in Section~\ref{sec:concl}.

\boldmath
\section{CP Violation}\label{sec:CPV}
\unboldmath
Since $B^0_s$ and $\bar B^0_s$ mesons may both decay into the same final state $D_s^{+} K^-$, we obtain interference between the corresponding decay amplitudes through $B^0_s$--$\bar B^0_s$ mixing, which leads to time-dependent decay rates. In order to probe the corresponding CP-violating effects, rate asymmetries of the following kind are considered  \cite{RF-BsDsK,DeBFKMST}:
\begin{equation}\label{CP-asym}
	 \frac{\Gamma(B^0_s(t)\to D_s^{+} K^-) - \Gamma(\bar{B}^0_s(t)\to D_s^{+} K^-) }
	{\Gamma(B^0_s(t)\to D_s^{+} K^-) + \Gamma(\bar{B}^0_s(t)\to D_s^{+} K^-) }  
	 = \frac{{C}\,\cos(\Delta M_s\,t) + {S}\,\sin(\Delta M_s\,t)}
	{\cosh(y_s\,t/\tau_{B_s}) + {\cal A}_{\Delta\Gamma}\,\sinh(y_s\,t/\tau_{B_s})};
\end{equation}
an analogous expression holds for the CP-conjugate final state $D_s^{-} K^+$, where $C$, $S$ and ${\cal A}_{\Delta\Gamma}$ are 
replaced by $\overline{C}$, $ \overline{S}$ and $\overline{{\cal A}}_{\Delta\Gamma}$, respectively. These observables satisfy the
following sum rules:
\begin{equation}\label{SR}
\Delta_{\rm SR}  \equiv   1-C^2-S^2-{\cal A}_{\Delta\Gamma}^2=0, 
\quad \overline{\Delta}_{\rm SR}  \equiv  1-\overline{C}^2-\overline{S}^2-\overline{{\cal A}}_{\Delta\Gamma}^2=0.
\end{equation}
In Eq.~(\ref{CP-asym}), $\Delta M_s\equiv M_{\rm H}^{(s)}-M_{\rm L}^{(s)}$ 
describes the mass difference of the $B_s$ mass eigenstates while their decay width 
difference $\Delta\Gamma_s\equiv \Gamma_{\rm L}^{(s)}-\Gamma_{\rm H}^{(s)}$ enters the parameter
\begin{equation}\label{ys}
y_s\equiv \frac{\Delta\Gamma_s}{2\,\Gamma_s}=0.062 \pm 0.004,
\end{equation}
where $\Gamma_s\equiv \tau_{B_s}^{-1}$ is the inverse of the average lifetime of the $B_s$ system, and the numerical value corresponds to the current experimental average \cite{PDG}. It should be noted that the decay width parameter $y_s$ is sizeable in the $B_s$ system, while its $B_d$ counterpart takes a value at the per mille level. Thanks to this feature, we may get experimental access to the observable ${\cal A}_{\Delta\Gamma}$ in $B_s$ decays, as can be seen in Eq.~(\ref{CP-asym}). 

The interference effects between $B^0_s$--$\bar{B}^0_s$ mixing and decay processes giving rise to the structure of Eq.~(\ref{CP-asym}) are described by the following observable \cite{RF-BsDsK}:
\begin{equation}  \label{xi-def}
    \xi=- e^{-i \phi_s}\left[ e^{i \phi_{\rm CP}(B_s)} \frac{A(\overline{B}^0_s 
    \rightarrow D_s^{+} K^{-})}{A(B^0_s \rightarrow D_s^{+} K^{-})} \right],
  \end{equation}
where the decay amplitudes $A(\overline{B}^0_s \rightarrow D_s^{+} K^{-})$ and $A(B^0_s \rightarrow D_s^{+} K^{-})$ 
enter and $\phi_s$ is the CP-violating $B^0_s$--$\bar{B}^0_s$ mixing phase. The latter quantity can be determined through 
CP-violating effects in $B^0_s\to J/\psi \phi$ and similar modes \cite{DDF,Dunietz:2000cr,Faller:2008gt,DeBF,Barel:2020jvf}. 
Using an average of the corresponding measurements and including 
corrections from doubly Cabibbo-suppressed penguin topologies through control channels 
yields 
\begin{equation}\label{phis}
\phi_s=\left(-5^{+1.6}_{-1.5}\right)^\circ,
\end{equation}
while the SM prediction takes a value around $-2^\circ$, as discussed in detail in Ref.~\cite{Barel:2020jvf}.

As discussed in detail in Ref.~\cite{RF-BsDsK}, the convention-dependent phase $\phi_{\rm CP}(B_s)$ in Eq.~(\ref{xi-def}), 
which is defined through the CP transformation
\begin{equation}
({\cal CP})|B^0_s\rangle=e^{-i\phi_{\rm CP}(B_s)}|\overline{B}^0_s\rangle,
\end{equation}
is cancelled by the ratio of the decay amplitudes $A(\overline{B}^0_s\to D_s^+K^-)$ and
$A({B}^0_s\to D_s^+K^-)$, where a CP transformation is required in the corresponding hadronic matrix element of the current--current operators to transform the $|B^0_s\langle$ into a $|\overline{B}^0_s\rangle$ state, yielding
\begin{equation}\label{xi-ME}
  \xi = - e^{-i(\phi_s + \gamma)} \left[ \frac{1}{x_s e^{i \delta_s}} \right].
\end{equation}
Here $x_s$ with the CP-conserving strong phase $\delta_s$ parametrizes the ratio of decay 
amplitudes, which depend on the corresponding hadronic matrix elements and elements of the 
Cabibbo--Kobayashi--Maskawa (CKM) matrix \cite{RF-BsDsK,DeBFKMST}. 
The interference effects of the decays into the CP-conjugate final state 
$D^-K^+$ are described by the observable 
\begin{equation}\label{xi-bar-ME}
 {\bar{\xi}} = - e^{-i(\phi_s + \gamma)} \left[{x_s e^{i \delta_s}} \right].
\end{equation}
We notice the following relation:
\begin{equation} \label{multxi}
 {\xi} \times \bar{\xi}= e^{-i2( \phi_s + \gamma)},
\end{equation} 
where the non-perturbative parameter $x_s e^{i \delta_s}$ cancels. Consequently, Eq.~(\ref{multxi}) is not affected by hadronic uncertainties, thereby representing a {\it theoretically clean} expression which offers access to the CP-violating phase
$ \phi_s + \gamma$. Using the value of $\phi_s$ in Eq.~(\ref{phis}), we may extract the UT angle $\gamma$.

The observables entering the time-dependent rate asymmetry in Eq.~(\ref{CP-asym}) are given in terms of $ \xi$ as 
follows:
\begin{equation}\label{obs-xi}
 C=\frac{1-|\xi|^2}{1+|\xi|^2},  \quad S= \frac{2\,\text{Im}{\,\xi}}{1 + |\xi|^2}, \quad 
 \mathcal{A}_{\Delta \Gamma}=\frac{2\,\text{Re}\,\xi}{1+|\xi|^2}\ .
\end{equation}
Similar expressions hold for the CP-conjugate observables, where $\xi$ is replaced by $ \bar{\xi}$. Note that 
these expressions actually satisfy the sum rules in Eq.\ (\ref{SR}).

It should be emphasised that Eqs.\ (\ref{xi-ME}) and (\ref{xi-bar-ME}) rely on the SM structure of the corresponding decay amplitudes, yielding
\begin{equation}\label{xi-rel}
| \bar{\xi}|=\frac{1}{|\xi|},
\end{equation}
which implies
\begin{equation}\label{C-rel}
 C+\overline{C}=0.
 \end{equation}
In the analysis presented in Ref.~\cite{LHCb-BsDsK}, the LHCb collaboration has assumed these relations.
 
 The absolute value of $\xi$ can be determined from the measured value of $C$ through
\begin{equation}\label{xi-det}
 |\xi|=\sqrt{\frac{1-C}{1+C}},
\end{equation} 
while further information from $\mathcal{A}_{\Delta \Gamma}$ and $S$ allows the extraction of the real and imaginary parts,
respectively, with the help of the expressions
\begin{equation}
\text{Re}\,\xi=\frac{{\cal A}_{\Delta\Gamma}}{1+C}, \quad \text{Im}\,\xi=\frac{S}{1+C},
\end{equation}
thereby fixing the complex $\xi$ from the data.
In analogy, the observables corresponding to the CP-conjugate state $D_s^-K^+$ allow us to determine $ \bar{\xi}$. We may then use the relation in Eq.\ (\ref{multxi}) to determine $ \phi_s + \gamma$ in a theoretically clean way. Due to the multiplicative factor of two associated with this phase, we obtain a twofold ambiguity, modulo $180^\circ$. 

In case we have only measurements of $C$ and $S$ available, we would obtain a twofold ambiguity for $\xi$, and in analogy for $ \bar{\xi}$. Consequently, we would then have a fourfold ambiguity for $ 2(\phi_s + \gamma)$ when applying Eq.~(\ref{multxi}), resulting in an eightfold ambiguity for $\phi_s + \gamma$, and finally $\gamma$ itself. However, due to the sizeable decay width difference 
$\Delta\Gamma_s$, we actually obtain access to the observables $ \mathcal{A}_{\Delta \Gamma}$ and 
$\overline{{\cal A}}_{\Delta\Gamma}$, thereby just leaving a twofold ambiguity, as was also pointed out 
in Refs.~\cite{RF-BsDsK,DeBFKMST}.

\begin{table}[t!]
    \centering
    \begin{tabular}{cc}
    \hline
   \hspace{4cm} Observables\\
   \hline
   \hline
 ${C}=-0.73 \pm 0.15$&  $\overline{C}=+0.73 \pm 0.15$\\
       $S=+0.49 \pm 0.21$  & $\overline{S}=+0.52 \pm 0.21 $\\
       $\mathcal{A}_{\Delta \Gamma}=+0.31 \pm 0.32$  & $\mathcal{\overline{A}}_{\Delta \Gamma}=+0.39 \pm 0.32$\\
       \hline 
        $\langle S \rangle_+= 0.50 \pm 0.15$ &  $\langle S \rangle_-= 0.02 \pm 0.15$\\
         $\langle \mathcal{A}_{\Delta \Gamma} \rangle_+= 0.35 \pm 0.23$ &  
         $\langle \mathcal{A}_{\Delta \Gamma} \rangle_-= 0.04 \pm 0.23$ \\
         \hline
       \end{tabular}
    \caption{CP-violating $B^0_s\to D_s^\mp K^\pm$ 
    observables corresponding to the LHCb analysis in Ref.~\cite{LHCb-BsDsK}, with the observable combinations introduced in 
    Eq.\ (\ref{obs-comb}). Note that the results assume the relation in Eq.\ (\ref{C-rel}).}
    \label{tab:expxixibar}
\end{table}

Using the LHCb results reported in Ref.\ \cite{LHCb-BsDsK} and taking the proper sign conventions into account, 
we obtain the observables collected in Table \ref{tab:expxixibar}, where we have added the statistical and systematic uncertainties
in quadrature. It is interesting to observe that these measured values are consistent with the sum rules in 
Eq.\ (\ref{SR}):
\begin{equation}
{\Delta}_{\rm SR}=0.13 \pm 0.36, \quad \overline{\Delta}_{\rm SR}=0.04\pm 0.40.
\end{equation}
We give also the values for the combinations
\begin{equation}\label{obs-comb}
\langle S \rangle_\pm\equiv \frac{\overline{S}\pm S}{2}, \quad   \langle \mathcal{A}_{\Delta \Gamma} \rangle_\pm\equiv
\frac{\mathcal{\overline{A}}_{\Delta \Gamma}\pm \mathcal{A}_{\Delta \Gamma}}{2},
\end{equation}
which will be useful below. Applying Eqs.\ (\ref{xi-rel}) and (\ref{xi-det}), we obtain
\begin{equation}\label{xi-det-2}
|\xi|=2.53^{+1.43}_{-0.59}, \quad |\bar{\xi}|=0.40 \pm 0.13.
\end{equation}

\begin{figure}[t]
	\centering
	\includegraphics[width = 0.49\linewidth]{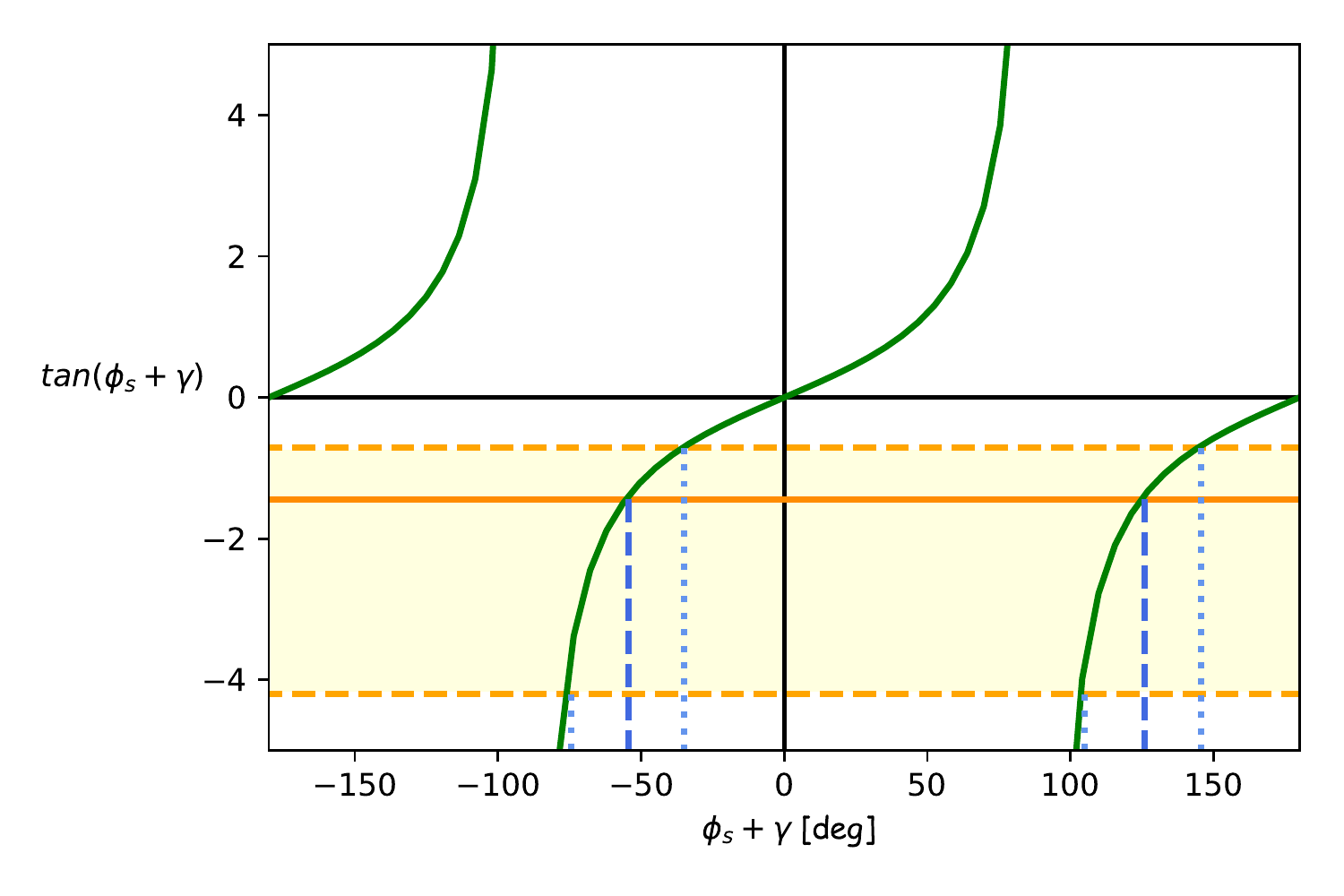}
	\includegraphics[width = 0.49\linewidth]{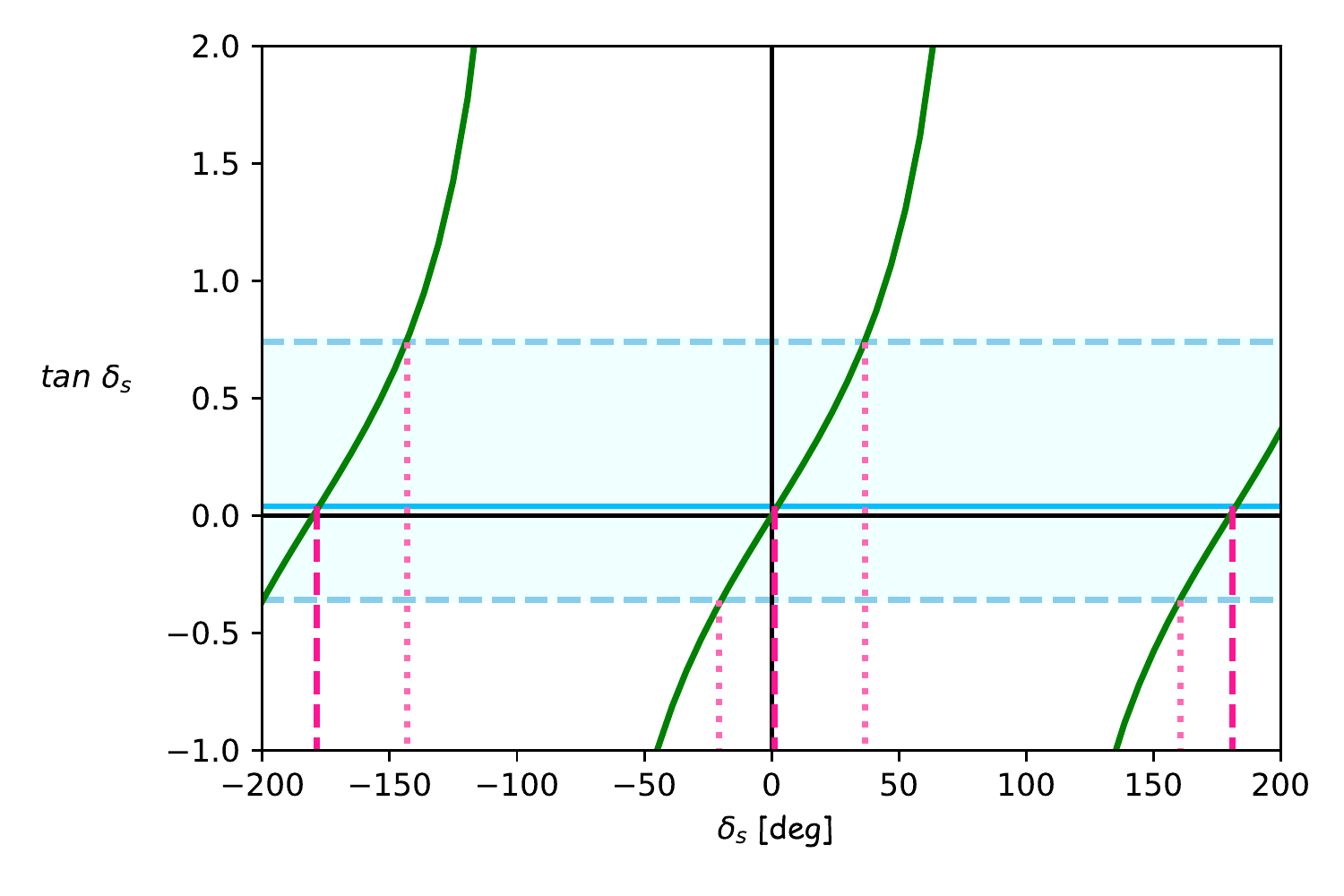}
	\caption{Illustration of $\tan(\phi_s+\gamma)$ (left) and $\tan\delta_s$ (right) 
	with their experimental values as given in Eqs.~(\ref{tanPhi}) and (\ref{tan-del}), respectively.}
		\label{fig:tan}
\end{figure}

In order to determine $\phi_s+\gamma$ and $\delta_s$, we may use the following relations \cite{RF-BsDsK,DeBFKMST}:
\begin{equation}\label{tanPhi}
\tan(\phi_s+\gamma)=-\frac{\langle S \rangle_+}{ \langle \mathcal{A}_{\Delta \Gamma} \rangle_+}=-1.45^{+0.73}_{-2.76}
\end{equation}
\begin{equation}\label{tan-del}
\tan\delta_s=\frac{\langle S \rangle_-}{ \langle \mathcal{A}_{\Delta \Gamma} \rangle_+}=0.04^{+0.70}_{-0.40},
\end{equation}
which we have illustrated in Fig.~\ref{fig:tan}. We note that the quantities $\langle S \rangle_{\pm}$ and $\langle\mathcal{A}_{\Delta \Gamma} \rangle_{\pm}$ have been calculated without taking correlations into account. In the plot in the panel on the left-hand side, we can nicely 
see how the twofold solution for $\phi_s+\gamma$ arises from the measured observables:
\begin{equation}\label{phis-gam-res}
\phi_s +\gamma= (-55^{+18}_{-22})^\circ \quad \lor \quad (125^{+18}_{-22})^\circ.
\end{equation}
Concerning the CP-conserving strong phase $\delta_s$, we obtain
\begin{equation}\label{del-res}
\delta_s= (182^{+34}_{-22})^\circ \quad \lor \quad (2^{+34}_{-22})^\circ,
\end{equation}
as illustrated in the plot in the panel on the right-hand side in Fig.~\ref{fig:tan}. Using 
\begin{equation}
\frac{\langle S \rangle_+}{\sqrt{1-C^2}}=+\cos\delta_s\sin(\phi_s+\gamma), \quad
\frac{\langle\mathcal{A}_{\Delta \Gamma} \rangle_+}{\sqrt{1-C^2}}=-\cos\delta_s\cos(\phi_s+\gamma)
\end{equation}
and taking the signs of $\langle S \rangle_+$ and $\langle\mathcal{A}_{\Delta \Gamma} \rangle_+$ in Table~\ref{tab:expxixibar}
into account, we observe that $(\phi_s+\gamma)\sim -55^\circ$ and $125^\circ$ are associated with $\delta_s \sim 180^\circ $
and $0 ^\circ$, respectively. Note that $\langle S \rangle_-$ and $\langle\mathcal{A}_{\Delta \Gamma} \rangle_-$ are both proportional to $\sin\delta_s$, which is also reflected by their small experimental values.  As pointed out in Ref.~\cite{RF-BsDsK}, the
case of $\delta_s\sim0 ^\circ$ corresponds to the picture of factorisation, which we will discuss in more detail in Subsection~\ref{ssec:fact}. Consequently, this framework allows us to single out the solution $(\phi_s+\gamma)\sim 125^\circ$ with
$\delta_s\sim0 ^\circ$, thereby excluding the values modulo $180^\circ$.

The LHCb collaboration has obtained results in Ref.~\cite{LHCb-BsDsK} that are consistent with our findings in 
Eqs.~(\ref{phis-gam-res}) and (\ref{del-res}). Performing a fit to the experimental data, taking also the relevant
correlations into account, a sharper picture arises:
\begin{equation}\label{LHCb-par-res}
\phi_s+\gamma=\left(126^{+17}_{-22}\right)^\circ, \quad \delta_s= (-2^{+13}_{-14})^\circ, \quad x_s=|\bar{\xi}|=0.37^{+0.10}_{-0.09}.
\end{equation}
Here we have used $\phi_s=(-1.7\pm1.9)^\circ$ to convert the value of $\gamma$ in Eq.~(\ref{LHCb-gam}) into $\phi_s+\gamma$, and have omitted the excluded solutions modulo $180^\circ$. Using the result for $\phi_s$ in Eq.\ (\ref{phis}), which takes 
penguin corrections into account, we obtain
\begin{equation}\label{gamma-res-1}
\gamma=\left(131^{+17}_{-22}\right)^\circ.
\end{equation}
The value for $|\bar{\xi}|$ is consistent with the one in Eq.\ (\ref{xi-det-2}).
 In Refs.~\cite{RF-BsDsK,DeBFKMST}, assuming the SM expressions
 for the relevant decay amplitudes, the hadronic parameters $x_s$ and $\delta_s$ were determined through data for $B^0_d \rightarrow D^{\pm}\pi^{\mp}$ decays, which are related to the $B^0_s\to D_s^\mp K^\pm$ system through the $U$-spin symmetry of strong interactions. These results are in good agreement with those in 
 Eq.\ (\ref{LHCb-par-res}) within the uncertainties. 

The result for $\gamma$ in Eq.~(\ref{gamma-res-1}) with its central value much larger than the regime of $70^\circ$, which arises from analyses of pure tree-level decays of the kind $B\to D K$ as well as from global fits of the UT within the SM, 
is intriguing despite its significant uncertainty. We note that this regime is also consistent with a recent simultaneous analysis of various tree decays of $B$ mesons to extract $\gamma$ as well as charm mixing and other hadronic parameters \cite{LHCb:2021dcr}. This analysis includes channels with decay dynamics different from the $B^0_s\to D_s^\mp K^\pm$ and related modes considered in our study. Moreover, the sensitivity on $\gamma$ arises from very different interference effects, while in our case of the $B^0_s\to D_s^\mp K^\pm$ system mixing-induced CP violation plays the central role. 

Within the SM, the determination of the value of $\gamma$ given in Eq.\ (\ref{gamma-res-1}) is theoretically clean and -- in particular -- not affected by 
strong-interactions effects. It shows a discrepancy with the results of $\gamma$ in the $70^\circ$ regime at the $3\,\sigma$ level. We need to shed more light on this puzzling situation, which would require new sources of CP violation. In principle, such effects could enter through $B^0_s$--$\bar B^0_s$ mixing. However, using the value of $\phi_s$ determined through experimental data, such effects are included. Thus, we would need  CP-violating NP contributions arising directly at the decay amplitude level, that should then also manifest themselves in the corresponding branching ratios. Therefore, let us next have a closer look at these quantities.

\boldmath
\section{Branching Ratios}\label{sec:BR}
\unboldmath
\subsection{Disentangling the Branching Ratios}
In the $B^0_s\to D_s^\mp K^\pm$ system, complications arise due to the interference between the different decay paths as well as the impact of $B^0_s$--$\bar B^0_s$ mixing. Let us first ``switch off" the mixing effects and focus on disentangling the different decay contributions, considering $\bar B^0_s$ and $B^0_s$ decays into the final state $D_s^{+} K^-$. We introduce the branching ratio
\begin{equation}\label{BR-th-expr}
\hspace*{-0.5truecm}\mathcal{B}_{\text{th}} \equiv \frac{1}{2}\left[ \mathcal{B}(\bar{B}^0_s \rightarrow D_s^+ K^-)_{\text{th}} +  
\mathcal{B}({B}^0_s \rightarrow D_s^+ K^-)_{\text{th}}\right],
\end{equation}
where the factor of one half arises from the average of the $\bar{B}^0_s$ and $B^0_s$ decays. The individual branching ratios
\begin{equation}\label{BR-theo-single}
 \mathcal{B}(\bar{B}^0_s \rightarrow D_s^+ K^-)_{\text{th}}=
|A(\bar{B}^0_s \rightarrow D_s^+ K^- )|^2  \, \Phi_{\rm Ph} \,\tau_{B_s} 
\end{equation}
\begin{equation}
 \mathcal{B}(B^0_s \rightarrow D_s^+ K^-)_{\text{th}}=
 |A(B^0_s \rightarrow D_s^+ K^- )|^2 \, \Phi_{\rm Ph} \,\tau_{B_s} 
\end{equation}
involve the corresponding decay amplitudes with the phase-space factor
\begin{equation}
\Phi_{\rm Ph}\equiv\frac{1}{16 \, \pi \, m_{B_s}} \, \Phi\left(\frac{m_{D_s}}{m_{B_s}},\frac{m_K}{m_{B_s}}\right),
\end{equation}
where the meson masses $m_{B_s}$, $m_{D_s}$ and $m_K$ enter the phase-space function
\begin{equation}\label{Phi-def}
\Phi(x,y) \equiv \sqrt{[1-(x+y)^2][1-(x-y)^2]}.
\end{equation}
The ``theoretical" branching ratio (\ref{BR-th-expr}) is related to the ``experimental" branching ratio
\begin{equation}
\mathcal{B}_{\text{exp}} = 
 \frac{1}{2} \int_0^{\infty} \! \left[\Gamma (\bar{B}_s^0 (t)\rightarrow D_s^+ K^-) + \Gamma (B_s^0 (t)\rightarrow D_s^+ K^-) \right]
 \mathrm{d}t,
\end{equation}
which corresponds to the time-integrated untagged decay rate \cite{Dunietz:2000cr}, as follows \cite{DeBruyn:2012wj}:
\begin{equation}\label{BR-dictionary}
\mathcal{B}_{\text{th}} = \left[ \frac{1-y_s^2}{1+ \mathcal{A}_{\Delta \Gamma_s} y_s} \right] \mathcal{B}_{\text{exp}}.
\end{equation}
Using Eq.~(\ref{xi-def}), we may write
\begin{equation}
\mathcal{B}_{\text{th}} = \frac{1}{2}\left(1+|\xi|^2\right)\mathcal{B}_{\text{th}}(B^0_s\to D_s^+K^-) =
\frac{1}{2}\left(1+|\xi|^{-2}\right)\mathcal{B}_{\text{th}}(\bar B^0_s\to D_s^+K^-),
\end{equation}
and obtain
\begin{equation}\label{BRbar-Ds+K-}
\mathcal{B}(\bar B^0_s\to D_s^+K^-)_{\text{th}}=2 \left(\frac{|\xi|^2}{1+|\xi|^2} \right)\mathcal{B}_{\text{th}}
\end{equation}
\begin{equation}\label{BR-Ds+K-}
\mathcal{B}(B^0_s\to D_s^+K^-)_{\text{th}}=2 \left(\frac{1}{1+|\xi|^2} \right)\mathcal{B}_{\text{th}},
\end{equation}
where $\mathcal{B}_{\text{th}}$ can be determined from the experimental branching ratio through Eq.~(\ref{BR-dictionary}).

For the $\bar{B}^0_s$ and $B^0_s$ 
decays into the CP-conjugate final state $D_s^- K^+$, we obtain analogous expressions, where we have to replace
$\mathcal{B}_{\text{exp}}$, $\mathcal{B}_{\text{th}}$, $\mathcal{A}_{\Delta \Gamma_s}$ and $\xi$ through their counterparts
$\bar{\mathcal{B}}_{\text{exp}}$, $\bar{\mathcal{B}}_{\text{th}}$, $\bar{\mathcal{A}}_{\Delta \Gamma_s}$ and $\bar{\xi}$, 
respectively. 
In analogy to Eqs.~(\ref{BRbar-Ds+K-}) and (\ref{BR-Ds+K-}), we may then also determine the theoretical branching ratios of the
$\bar B^0_s\to D_s^-K^+$ and $B^0_s\to D_s^-K^+$ modes. In the SM, we have the following relations:
\begin{equation}\label{SM-BR-1}
\mathcal{B}(\bar B^0_s\to D_s^+K^-)_{\text{th}}\stackrel{\rm SM}{=}\mathcal{B}(B^0_s\to D_s^-K^+)_{\text{th}}
\end{equation}
\begin{equation}
\mathcal{B}(B^0_s\to D_s^+K^-)_{\text{th}}\stackrel{\rm SM}{=}\mathcal{B}(\bar B^0_s\to D_s^-K^+)_{\text{th}},
\end{equation}
yielding
\begin{equation}
\mathcal{B}_{\text{th}}\stackrel{\rm SM}{=}\bar{\mathcal{B}}_{\text{th}}.
\end{equation}

\begin{table}[b!]
    \centering
    \begin{tabular}{c|c}
    \hline
      \hspace{1.5cm} Parameters   \hspace{1.5cm} &       \hspace{1.5cm}  Values       \hspace{1.5cm} \\
    \hline
    \hline
           $  \mathcal{B}_{\rm th}$ & $(1.10 \pm 0.09) \times 10^{-4}$\\
        $  \mathcal{B}(\bar{B}^0_s \rightarrow D_s^{+}K^{-})_{\rm th}$ & $(1.94 \pm 0.21) \times 10^{-4}$\\
          $  \mathcal{B}(B^0_s \rightarrow D_s^{+}K^{-})_{\rm th}$  & $(0.26 \pm 0.12) \times 10^{-4}$ \\
        \hline
    \end{tabular}
    \caption{Values of the theoretical branching ratios characterising the $B^0_s\to D_s^\mp K^\pm$ system, assuming vanishing NP contributions to the corresponding decay amplitudes, as discussed in the text.}
    \label{tab:xi_BRth}
\end{table}

It would be very interesting to test these SM relations through separate measurements of the experimental 
branching ratios $\mathcal{B}_{\text{exp}}$ and $\bar{\mathcal{B}}_{\text{exp}}$. Unfortunately, such results 
have not yet been reported. However, there are measurements of the following average \cite{PDG}:
\begin{equation}\label{SM-BR-2}
\mathcal{B}^{\text{exp}}_\Sigma \equiv \mathcal{B}_{\text{exp}} + \bar{\mathcal{B}}_{\text{exp}} = (2.27 \pm 0.19) \times 10^{-4}.
\end{equation}
Assuming now, as was also done by the LHCb collaboration \cite{LHCb-BsDsK}, again the framework of the SM, 
we obtain the following relation \cite{DeBFKMST}:
\begin{equation}\label{SM-BR-3}
 \mathcal{B}_{\text{th}} =  \bar{\mathcal{B}}_{\text{th}}=
\left[\frac{1-y_s^2}{1+y_s\langle {\cal A}_{\Delta\Gamma}\rangle_+}\right]\langle\mathcal{B}_{\text{exp}}\rangle,
\end{equation}
where $\langle \mathcal{A}_{\Delta \Gamma} \rangle _{+}$ was introduced in Eq.~(\ref{obs-comb}), and
\begin{equation}
\langle\mathcal{B}_{\text{exp}}\rangle\equiv \frac{1}{2}\left(\mathcal{B}_{\text{exp}} + \bar{\mathcal{B}}_{\text{exp}}\right)=
\frac{1}{2} \, \mathcal{B}^{\text{exp}}_\Sigma.
\end{equation}
Using then Eqs.~(\ref{BRbar-Ds+K-}) and (\ref{BR-Ds+K-}) and the value of $|\xi|$ extracted from the experimental study of CP violation discussed in Section~\ref{sec:CPV}, we may determine the
theoretical branching ratios of the $\bar B^0_s\to D_s^+K^-$ and $B^0_s\to D_s^+K^-$ decays, which are -- due to our assumption of the SM -- equal to their CP conjugates. In Table~\ref{tab:xi_BRth}, we summarise the corresponding results following from 
the current data.

\subsection{Consistency of the Branching Ratios with Partner Decays and First Information on Exchange Topologies}
It is instructive to compare the results collected in Table~\ref{tab:xi_BRth} with the branching ratios 
of the $\bar{B}^0_d\to D_d^+K^-$ and $B^0_d\to D_s^+\pi^-$ decays, which 
differ from the $\bar{B}^0_s\to D_s^+K^-$ and $B^0_s\to D_s^+K^-$ modes, respectively, 
only through their spectator quarks and originate from
the same quark-level processes. The current experimental results, which are actually CP-averaged branching ratios, 
read as follows \cite{PDG}:
\begin{equation}\label{BR-PDG-1}
 \mathcal{B}(\bar{B}^0_d\to D_d^+K^-)=(1.86\pm0.20)\times10^{-4}, \quad
 \mathcal{B}({B}^0_d\to D_s^+\pi^-)=(2.16\pm0.26)\times10^{-5}.
\end{equation}
In contrast to their $B_s$ counterparts, these modes do not receive contributions from exchange topologies. We may use the branching ratios to determine the ratios 
\begin{equation}\label{E-T-1}
\left|\frac{T_{D_sK}}{T_{D_dK}}\right|^2\left|1+\frac{E_{D_sK}}{T_{D_sK}} \right|^2=\frac{\tau_{B_d}}{\tau_{B_s}}\frac{m_{B_s}}{m_{B_d}}\left[\frac{\Phi(m_{D_d}/m_{B_d}, m_{K}/m_{B_d})}{\Phi(m_{D_s}/m_{B_s}, m_{K}/m_{B_s})}\right]
\left[\frac{\mathcal{B}(\bar{B}^0_s \rightarrow D_s^{+}K^{-})_{\rm th}}{\mathcal{B}(\bar{B}^0_d\to D_d^+K^-)}
\right]
\end{equation}
\begin{equation}\label{E-T-2}
\left|\frac{T_{KD_s}}{T_{\pi D_s}}\right|^2\left|1+\frac{E_{KD_s}}{T_{KD_s}} \right|^2=\frac{\tau_{B_d}}{\tau_{B_s}}\frac{m_{B_s}}{m_{B_d}}\left[\frac{\Phi(m_{D_s}/m_{B_d}, m_{\pi}/m_{B_d})}{\Phi(m_{D_s}/m_{B_s}, m_{K}/m_{B_s})}\right]
\left[\frac{\mathcal{B}({B}^0_s \rightarrow D_s^{+}K^-)_{\rm th}}{\mathcal{B}({B}^0_d\to D_s^+\pi^-)}
\right],
\end{equation}
where the $E$ and $T$ amplitudes describe the corresponding exchange and colour-allowed tree topologies. The $SU(3)$ flavour symmetry of strong interactions implies
\begin{equation}\label{T-ampls}
T_{D_sK} \approx T_{D_dK}, \quad T_{KD_s} \approx T_{\pi D_s},
\end{equation}
where $SU(3)$-breaking corrections may only arise from the spectator quarks. Using the 
experimental results in Table~\ref{tab:xi_BRth} and Eq.~(\ref{BR-PDG-1}) with the meson masses in Table~\ref{masses} and 
the following values of the average lifetimes of the $B^0_s$  and $B^0_d$ mesons \cite{PDG}:
\begin{equation}\label{lifetimes}
\tau_{B_s}=(1.527 \pm 0.011) \,{\rm ps}, \quad \tau_{B_d}=(1.519\pm0.004)\,{\rm ps}, 
\end{equation}
we obtain 
\begin{equation}\label{E-T-1-res}
\left|\frac{T_{D_sK}}{T_{D_dK}}\right|\left|1+\frac{E_{D_sK}}{T_{D_s K}} \right|=1.03 \pm 0.08
\end{equation}
\begin{equation}\label{E-T-2-res}
\left|\frac{T_{KD_s}}{T_{\pi D_s}}\right| \left|1+\frac{E_{KD_s}}{T_{K D_s}} \right|=1.11 \pm0.26.
\end{equation}
These findings are consistent with a smallish impact of the exchange topologies, which was also found in 
Refs.~\cite{DeBFKMST,FST-BR}. We shall quantify this picture in more detail in our analysis in Subsection~\ref{ssec:fact}.

\begin{table}[t!]
    \centering
    \begin{tabular}{c|c}
    \hline
 \hspace{1cm} Masses  \hspace{1cm} &  \hspace{1.5cm} Values  \hspace{1.5cm} \\
   \hline
   \hline
       $m_{B_s}$  & $(5366.88 \pm 0.14) {\text{\ MeV}}$  \\   
        $m_{B_d}$ & $(5279.64 \pm 0.12) {\text{\ MeV}}$ \\      
       $m_{D_s}$ &  $(1968.34 \pm 0.07) {\text{\ MeV}}$ \\        
       $m_{D_d}$ &  $(1869.66 \pm 0.05) {\text{\ MeV}}$ \\   
       $m_{K}$ & $(493.677 \pm 0.016) {\text{\ MeV}}$  \\
       $m_{\pi}$ & $(139.5704 \pm 0.0002) {\text{\ MeV}}$ \\
             \hline          
    \end{tabular}
       \caption{Meson masses relevant for our numerical analysis \cite{PDG}.} \label{masses} 
\end{table}

The decays $\bar{B}^0_d\to D_d^+\pi^-$ and $\bar{B}^0_s\to D_s^+\pi^-$ originate from $b\to c \bar u d$ quark-level processes and
are related to the $\bar{B}^0_s\to D_s^+K^-$ and $\bar{B}^0_d\to D_d^+K^-$ modes through the $U$-spin symmetry of strong interactions, respectively \cite{RF-BsDsK,DeBFKMST}, allowing us to determine 
\begin{equation}\label{E-T-3}
\left|\frac{T_{D_d \pi}}{T_{D_s\pi}}\right|^2\left|1+\frac{E_{D_d \pi}}{T_{D_d \pi}} \right|^2=\frac{\tau_{B_s}}{\tau_{B_d}}\frac{m_{B_d}}{m_{B_s}}\left[\frac{\Phi(m_{D_s}/m_{B_s}, m_{\pi}/m_{B_s})}{\Phi(m_{D_d}/m_{B_d}, m_{\pi}/m_{B_d})}\right]
\left[\frac{\mathcal{B}(\bar{B}^0_d \rightarrow D_d^{+}\pi^{-})}{\mathcal{B}(\bar{B}^0_s\to D_s^+\pi^-)_{\rm th}}
\right].
\end{equation}
Using the relation 
\begin{equation} \label{Dspi_th}
\mathcal{B}(\bar{B}^0_s\to D_s^+\pi^-)_{\rm th}=(1-y_s^2)\,\mathcal{B}(\bar{B}^0_s\to D_s^+\pi^-)_{\rm exp}
\end{equation}
between the theoretical and experimental $\bar{B}^0_s\to D_s^+\pi^-$ branching ratios \cite{DeBruyn:2012wj}, and 
the following experimental results \cite{PDG}
\begin{equation}\label{BR-values}
\mathcal{B}(\bar{B}^0_d \rightarrow D_d^{+}\pi^{-})=(2.52\pm0.13)\times10^{-3}, 
\quad \mathcal{B}(\bar{B}^0_s\to D_s^+\pi^-)_{\rm exp}= (3.00\pm0.23)\times10^{-3},
\end{equation}
we obtain 
\begin{equation}\label{TDspi-rat}
\left|\frac{T_{D_d \pi}}{T_{D_s\pi}}\right|\left|1+\frac{E_{D_d \pi}}{T_{D_d \pi}} \right|=0.91\pm0.04,
\end{equation}
which is consistent with Eq.~(\ref{E-T-1-res}) within the uncertainties.

\boldmath
\subsection{Factorisation}\label{ssec:fact}
\unboldmath
The calculation of non-leptonic $B$-meson decays is challenging due to the impact of strong interactions. 
Since decades, the ``factorisation" approach is applied as a particularly useful tool, where 
the hadronic matrix elements of the corresponding four-quark operators are factorised into the matrix elements of their
quark currents. Factorisation is not a universal feature of non-leptonic $B$ decays. Key examples where it is expected to work 
very well is given by decays of the kind $\bar B^0_s\to D_s^{+}\pi^-$ and $\bar B^0_d\to D_d^{+}K^-$, which originate only from 
colour-allowed tree-diagram-like topologies \cite{Beneke:2000ry,bjor,DG,Neubert:1997uc,SCET}.

The $\bar B^0_s\to D_s^{+}K^-$ channel, which plays a key role for our analysis, differs from $\bar B^0_d\to D_d^{+}K^-$ only through the spectator quark and is also caused by $b\to c \bar u s$ quark-level transitions. As we have seen above, it receives an additional contribution from an exchange topology, which involves the spectator quark and does not factorise. However, experimental data show that such exchange topologies contribute to the decay amplitudes at the few-percent level \cite{FST-BR}, thereby playing a minor role. In Eq.\ (\ref{E-T-1-res}), we have obtained a result consistent with these findings, which are
actually also theoretically expected \cite{FST-BR}.

Within the SM, we may write the $\bar B^0_s\to D_s^{+}K^-$ decay amplitude as follows:
\begin{equation}\label{eq:a1DsK}
A_{\bar{B^0_s} \rightarrow D_s^+ K^-}^{\text{SM}} = \frac{G_{\rm F}}{\sqrt{2}} \ V_{us}^{\ast}V_{cb} \ f_K  \ F_0^{B_s \rightarrow D_s}(m_K^2) \  (m_{B_s}^2 - m_{D_s}^2)  \ a_{\rm 1 \, eff }^{D_s K},
\end{equation}
where $G_{\rm F}$ is the Fermi constant, $V_{us}^{\ast}V_{cb}$ contains the relevant CKM matrix elements, $f_K$ is the kaon decay constant, and $F_0^{B_s \rightarrow D_s}(m_K^2)$ is a form factor entering the parametrisation of the hadronic 
$b \rightarrow c$ quark-current matrix element:
\begin{eqnarray}
\lefteqn{\langle D_s^+(k)|\overline{c}\gamma_\mu b|\bar B^0_s(p)
\rangle=}\nonumber\\
&&=F_0(q^2)\left(\frac{m_{B_s}^2-m_{D_s}^2}{q^2}\right)q_\mu+
F_1(q^2)\left[(p+k)_\mu -\left(\frac{m_{B_s}^2-m_{D_s}^2}{q^2}\right)q_\mu\right],\label{FF-para}
\end{eqnarray}
where $q\equiv p-k$ denotes the four-momentum transfer. The form factors can be calculated with a variety of approaches, most notably lattice QCD \cite{Monahan:2017uby, McLean:2019qcx, Aoki:2019cca}.
The parameter 
\begin{equation}\label{a-eff-1-DsK}
a_{\rm 1 \, eff }^{D_s K}=a_{1}^{D_s K} \left(1+\frac{E_{D_s K}}{T_{D_s K}}\right)
\end{equation}
describes the deviation from naive factorisation. Here $a_{1}^{D_s K}$ characterises the 
non-factorisable effects entering the colour-allowed tree amplitude $T_{D_s K}$, while $E_{D_s K}$ describes the non-factorisable exchange topologies, as introduced in Eq.~(\ref{E-T-1}).

In the analysis within the QCD factorisation approach in Ref.~\cite{Beneke:2000ry}, the $a_1$ parameters for 
colour-allowed $\bar{B}\to D P$ decays ($P=\pi, K$) originating from $b\to c \bar u r$ quark-level transitions ($r=d,s$) are found 
as $|a_1|\approx 1.05$ with a quasi-universal behaviour, which illustrates that factorisation is expected to work very well in this 
decay class. Another indication of this feature comes from the stable behaviour of $a_1$ under the QCD 
renormalisation group evolution  \cite{Buras:1994ij,Buras:1998us}, which is in contrast to the $a_2$ coefficient characterising 
colour-suppressed decays, where factorisation is not expected to work well. Interestinlgy, for decays of the kind $\bar{B}^0_d\to J/\psi \pi^0$, experimental data give values for $a_2(\bar{B}^0_d\to J/\psi \pi^0)$ that are surprisingly consistent with the picture of naive factorisation \cite{Barel:2020jvf}. 

The current state-of-the-art results for the decays  $\bar B^0_d\to D_d^{+}K^-$,  $\bar B^0_d\to D_d^{+}\pi^-$
and $\bar B^0_s\to D_s^{+}\pi^-$ calculated in QCD factorisation are given as follows \cite{Huber:2016xod, Bordone:2020gao}:
\begin{equation}\label{a1-pred0} 
|a_1^{D_dK}| = 1.0702^{+0.0101}_{-0.0128}\ , \quad
|a_1^{D_d\pi}| = 1.073^{+0.012}_{-0.014}\ , \quad
|a_1^{D_s\pi}| = 1.0727^{+0.0125}_{-0.0140} \ .
\end{equation}
In Ref.~\cite{Beneke:2021jhp}, even QED effects were studied, which are small and fully included within the uncertainties. 
These modes are related to one another through the $SU(3)$ flavour symmetry of strong interactions, leading to an essentially 
negligible difference of their $|a_1|$ parameters. We observe here also the quasi-universal behaviour noted above. Since the
$\bar{B}^0_s\to D_s^+K^-$ mode differs from the $\bar{B}^0_d\to D^+K^-$ channel only through the spectator quarks, we may
identify their $|a_1|$ parameters and shall use 
\begin{equation}\label{a1-1-pred}
|a_1^{D_sK}| = 1.07\pm0.02 ,
\end{equation}
 where we have doubled the tiny error in view of $SU(3)$-breaking effects in the spectator quarks, taking the spread of the $SU(3)$-related values in Eq.~(\ref{a1-pred0}) into account. 
 
 We may now calculate the ratios of the colour-allowed tree amplitudes entering the expressions in Eqs.~(\ref{E-T-1-res}) 
 and (\ref{TDspi-rat}):
 \begin{equation}
\left| \frac{T_{D_sK}}{T_{D_dK}}\right|=\left[\frac{F_0^{B_s \rightarrow D_s}(m_K^2)}{F_0^{B_d \rightarrow D_d}(m_K^2)}\right]\left[
\frac{m_{B_s}^2 - m_{D_s}^2}{ m_{B_d}^2 - m_{D_d}^2}\right]\left|\frac{a_{\rm 1}^{D_s K}}{a_{\rm 1}^{D_d K}}\right|=1.03\pm0.03
 \end{equation}
\begin{equation}
\left| \frac{T_{D_d\pi}}{T_{D_s\pi}}\right|=\left[\frac{F_0^{B_d \rightarrow D_d}(m_{\pi}^2)}{F_0^{B_s \rightarrow D_s}(m_{\pi}^2)}
\right]\left[\frac{m_{B_d}^2 - m_{D_d}^2}{ m_{B_s}^2 - m_{D_s}^2}\right]\left|\frac{a_{\rm 1}^{D \pi}}{a_{\rm 1}^{D_s \pi}}\right| =
0.99\pm0.03,
 \end{equation}
 where we have used the following theoretical form-factor ratios \cite{Bordone:2019guc}:
\begin{equation}
\left|\frac{F_0^{B_s \rightarrow D_s}(m_{\pi}^2)}{F_0^{B_d \rightarrow D_d}(m_{\pi}^2)}\right|=1.01\pm0.02,  \quad
\left|\frac{F_0^{B_s \rightarrow D_s}(m_K^2)}{F_0^{B_d \rightarrow D_d}(m_K^2)}\right|=1.01\pm0.02.
\end{equation}
Finally, using the numerical values in Eqs.~(\ref{E-T-1-res}) and (\ref{TDspi-rat}) following from experimental data, we obtain 
\begin{equation}\label{rE-values}
r_E^{D_sK}\equiv\left|1+\frac{E_{D_sK}}{T_{D_sK}} \right|=1.00\pm0.08, \quad 
r_E^{D_d\pi}\equiv\left|1+\frac{E_{D_d \pi}}{T_{D_d \pi}} \right|=0.92\pm0.05.
\end{equation}
These current state-of-the-art results do not indicate any anomalous behaviour of the exchange topologies, and are 
consistent with those obtained in Refs.~\cite{DeBFKMST,FST-BR}. 

Interestingly, we can also obtain direct insights into the importance of the exchange topologies. The decay $\bar{B^0_s} \to D^+ \pi^-$ arises only from such diagrams and is related to the exchange topology in $\bar{B^0_s} \to D_s^+ K^-$ through the $SU(3)$ flavour symmetry by replacing the $d\bar d$  through $s\bar s$ quark pairs. In analogy, the decay $\bar B^0_s\to D^-\pi^+$ is related to the exchange contribution to $\bar B^0_s\to D_s^-K^+$. Concerning the experimental status, there is only an upper bound $\mathcal{B}(B^0_s\to D^{*\mp} \pi^\pm)_{\rm exp}<6.1\times 10^{-6}$ (90\% C.L.) available \cite{Amhis:2019ckw}, while constraints on the branching ratios of the $\bar{B^0_s} \to D^+ \pi^-$ and $\bar{B^0_s} \to D^- \pi^+$ 
channels have not yet been reported. 

Another decay which originates only from an 
exchange topology is the $\bar{B^0_d} \to D_s^+ K^-$ mode, which differs from the exchange contribution to $\bar{B^0_s} \to D_s^+ K^-$ through the down quark of the initial $\bar{B^0_d}$ meson. Consequently, the $SU(3)$ flavour symmetry offers a relation between the corresponding amplitudes. This decay has actually been observed \cite{Amhis:2019ckw}:
\begin{equation}\label{E-BR-1}
\mathcal{B}(\bar{B^0_d} \to D_s^+ K^-)=(2.7\pm0.5)\times10^{-5}.
\end{equation}
If we employ
\begin{equation}
A(\bar{B^0_d} \to D_s^+ K^-)\equiv V_{cb}V_{ud}^* \, E_{D_sK}', \quad
A(\bar{B^0_s} \to D_s^+ K^-)\equiv V_{cb}V_{us}^* \, (T_{D_sK}+E_{D_sK}),
\end{equation}
we obtain
\begin{equation}
\hspace*{-0.2truecm}\left|\frac{E_{D_sK}'}{T_{D_sK}+E_{D_sK}}\right|=
\frac{\tau_{B_s}}{\tau_{B_d}}\frac{m_{B_d}}{m_{B_s}}
\left[\frac{\Phi(m_{D_s}/m_{B_s}, m_{K}/m_{B_s})}{\Phi(m_{D_s}/m_{B_d}, m_{K}/m_{B_d})}\right]
\left|\frac{V_{us}}{V_{ud}}\right|
\left[\frac{\mathcal{B}(\bar{B^0_d} \to D_s^+ K^-)}{\mathcal{B}(\bar{B^0_s} \to D_s^+ K^-)_{\rm th}}\right].
\end{equation}
Using then the results in Table~\ref{tab:xi_BRth} and Eq.\ (\ref{E-BR-1}) yields
\begin{equation}\label{E-rat-1}
\left|\frac{E_{D_sK}'}{T_{D_sK}+E_{D_sK}}\right| = 0.08 \pm 0.01,
\end{equation}
which is in excellent agreement with the picture in Eq.~(\ref{rE-values}). As was pointed out in Refs.~\cite{DeBFKMST,FST-BR}, the 
non-factorisable contributions to the exchange topologies suggest a large strong phase difference between the exchange 
and colour-allowed tree amplitudes, which is also supported by data for other modes. Consequently, we consider the 
ranges in Eq.~(\ref{rE-values}) as conservative assessments of the impact of the exchange topologies.

Let us now have a look at the $\bar{B^0_s} \rightarrow K^+ D_s^-$ decay, which originates from $b\to u \bar c s$ quark-level
transitions. In the SM, we may write the decay amplitude -- in analogy to Eq.~(\ref{eq:a1DsK}) -- in the form 
\begin{equation}
A_{\bar{B^0_s} \rightarrow K^+ D_s^-}^{\text{SM}} = \frac{G_{\rm F}}{\sqrt{2}} \ V_{cs}^{\ast}V_{ub} \  f_{D_s} \  F_0^{B_s \rightarrow K}(m_{D_s}^2) \ (m_{B_s}^2 - m_K^2) \ a_{\rm 1 \, eff }^{K D_s} 
\label{eq:a1KDs}
\end{equation}
with
\begin{equation}
a_{\rm 1 \, eff }^{K D_s}=a_{1}^{K D_s} \left(1+\frac{E_{K D_s}}{T_{K D_s}}\right),
\end{equation}
where the CKM factors are replaced correspondingly, $f_{D_s}$ is the $D_s$ decay constant, and 
$F_0^{B_s \rightarrow K}(m_{D_s}^2)$ parametrises the hadronic matrix element of the $b\to u$ transition. 
The coefficient $a_{1}^{K D_s}$ describes 
non-factorisable contributions to the colour-allowed tree amplitude $T_{K D_s}$, while the amplitude $E_{K D_s}$ arises from 
non-factorisable exchange topologies, as in Eq.\ (\ref{a-eff-1-DsK}). Although this channel is also colour-allowed (see the comments in the paragraph after Eq.~(\ref{a-eff-1-DsK})), the heavy-quark 
arguments for QCD factorisation for the $b\to c$ tree-level decays not apply in this case \cite{Beneke:2000ry}. As a reference point, we shall use the following range:
\begin{equation}\label{a1-2-pred}
|a_1^{K D_s}| = 1.1 \pm 0.1,
\end{equation} 
For the uncertainty, we use the picture arising from the QCD renormalisation group analysis in Ref.~\cite{Buras:1994ij} as guidance, 
where $a_1=1.01\pm0.02$ is found
for the global factorisation parameter $a_1$ for color-allowed decays, varying the QCD dimensional transmutation parameter and renormalisation scale within their allowed ranges. The corresponding uncertainty describes non-factorizable effects, as they have to cancel these dependences. Interestingly, the uncertainty is in excellent agreement with the state-of-the-art QCD factorisation results for the $b \to c $ modes in (\ref{a1-pred0}), where heavy-quark arguments can be used to prove factorisation up to tiny non-factorisable effects. Since these arguments do not apply in the $b\to u$ case, although we have still a colour-allowed decay, we will assume the uncertainty in (\ref{a1-2-pred}), which is five times larger, for our subsequent numerical studies. Interestingly, the strong phase difference $\delta_s$ in Eq.\ (\ref{LHCb-par-res}), with a central value close to $0^\circ$, disfavours large non-factorisable long-distance effects entering through the $\bar{B^0_s} \rightarrow K^+ D_s^-$ decay path.  It would be very important to put the $a_1^{K D_s}$ parameter on a solid theoretical basis in the future.

The amplitude of the decay $\bar{B}^0_d\to \pi^+D_s^-$ takes the same form as Eq.\ (\ref{eq:a1KDs}). However, this channel does not 
have contributions from exchange topologies, thereby yielding
\begin{equation}
a_{\rm 1 \, eff }^{\pi D_s}=a_{1}^{\pi D_s}. 
\end{equation}
The $\bar{B}^0_d\to \pi^+D_s^-$ channel differs from $\bar{B}^0_s\to K^+D_s^-$ through the spectator quarks, which are related through the $SU(3)$ flavour symmetry \cite{Gronau:1995hm}. Consequently, we assume 
\begin{equation}\label{a-1-rel-u}
|a_1^{\pi D_s}|=|a_1^{K D_s}| = 1.1 \pm 0.1,
\end{equation}
where we have used our reference value in Eq.\ (\ref{a1-2-pred}). Applying the formulae given above yields
\begin{equation}
\left|\frac{T_{KD_s}}{T_{\pi D_s}}\right|=
\left[\frac{F_0^{B_s \rightarrow K}(m_{D_s}^2)}{F_0^{B_d \rightarrow \pi}(m_{D_s}^2)}\right]\left[
\frac{m_{B_s}^2 - m_K^2}{ m_{B_d}^2 - m_{\pi}^2}\right]\left|\frac{a_{\rm 1}^{K D_s}}{a_{\rm 1}^{\pi D_s}}\right|. 
\end{equation}
The $SU(3)$-breaking effects in the $B_s \rightarrow K$ and $B_d \rightarrow \pi$ form factors were calculated with QCD light cone sum rules in Ref.~\cite{Khodjamirian:2017fxg}. Using these results, we assume the following numerical range in our analysis (see also the remark after Eq.~(\ref{FF-rat-0-1})):
\begin{equation}\label{FF-SU3-0}
\left[ \frac{F_0^{B_s \rightarrow K}(m_{D_s}^2)}{F_0^{B \rightarrow \pi}(m_{D_s}^2)} \right] =1.12 \pm 0.11.
\end{equation}
Finally, employing Eq.\ (\ref{a-1-rel-u}) and the meson masses in Table~\ref{masses}, we find 
\begin{equation}
\left|\frac{T_{KD_s}}{T_{\pi D_s}}\right|=1.15\pm0.19,
\end{equation}
which allows us to extract
\begin{equation}\label{E-range-1}
r_E^{KD_s}\equiv\left|1+\frac{E_{K D_s}}{T_{K D_s}} \right|=0.97 \pm 0.17
\end{equation}
from the numerical result in Eq.~(\ref{E-T-2-res}) following from the experimental data. We observe a pattern similar to the 
constraints in  Eq.\ (\ref{rE-values}), although with larger uncertainty. 

In analogy to the $\bar{B^0_d} \to D_s^+ K^-$ channel, the decay $\bar{B^0_d} \to D_s^- K^+$ is related to the exchange topology of the $\bar{B^0_s} \rightarrow K^+ D_s^-$ , differing only through the down quark of the initial $\bar{B^0_d}$ meson. Unfortunately, no measurement of the corresponding branching ratio is yet available \cite{Amhis:2019ckw}. Using 
\begin{equation}
\hspace*{-0.4truecm}\left|\frac{E_{D_sK}'}{T_{KD_s}+E_{KD_s}}\right|=
\frac{\tau_{B_s}}{\tau_{B_d}}\frac{m_{B_d}}{m_{B_s}}
\left[\frac{\Phi(m_{D_s}/m_{B_s}, m_{K}/m_{B_s})}{\Phi(m_{D_s}/m_{B_d}, m_{K}/m_{B_d})}\right]
\left|\frac{V_{ub}V_{cs}}{V_{cb}V_{ud}}\right|
\left[\frac{\mathcal{B}(\bar{B^0_d} \to D_s^- K^+)}{\mathcal{B}(\bar{B^0_s} \to K^+D_s^-)_{\rm th}}\right]
\end{equation}
with
\begin{equation}
\left|\frac{V_{ub}V_{cs}}{V_{cb}V_{ud}}\right|=\left[\frac{\lambda R_b}{1-\lambda^2/2}\right]\left[1+{\cal O}(\lambda^2)\right]=
0.089\pm0.005,
\end{equation}
where we have used the Wolfenstein parameterization and $R_b$ denotes the UT side from the origin to the apex \cite{CKMfitter}, 
we obtain
\begin{equation}
\left|\frac{E_{D_sK}'}{T_{KD_s}+E_{KD_s}}\right|=0.09\pm0.02
\end{equation}
from the experimental results in Table~\ref{tab:xi_BRth} and Eq.~(\ref{E-BR-1}). This numerical result is in excellent agreement with
Eq.~(\ref{E-rat-1}). Since the hadronic matrix elements of the exchange amplitudes scale with the product of the decay constants of the involved mesons, i.e.\ with $f_{B_{ds}}f_{D_s}f_K$ in the case of our modes, we obtain the relation $E_{D_sK}'\approx E_{K D_s}$. 
In view of the discussion after Eq. (\ref{E-rat-1}), we will consider the numerical range
\begin{equation}\label{rEKDs-range}
r_E^{KD_s}=1.00\pm0.08,
\end{equation}
which is similar as the one for $r_E^{D_sK}$ in Eq.\ (\ref{rE-values}), for our analysis of the 
$\bar{B^0_s} \rightarrow K^+ D_s^-$ decay path. This range is fully consistent with (\ref{E-range-1}), although giving a sharper picture. 

We would like to extract the $|a_1|$ parameters of the $\bar B^0_s\to D_s^{+}K^-$ and $\bar{B^0_s} \rightarrow K^+ D_s^-$  
channels from the data in the cleanest possible way, comparing them with the theoretical expectations. 
The central question is whether we will again encounter a puzzling situation as in Section~\ref{sec:CPV}.
In this respect, semileptonic decays provide a very useful tool.

\boldmath
\subsection{Information from Semileptonic Decays}\label{ssec:SL}
\unboldmath
\subsubsection{Preliminaries}
Using expressions (\ref{BR-th-expr})--(\ref{Phi-def}) and the decay amplitudes in Eqs.~(\ref{eq:a1DsK}) and (\ref{eq:a1KDs}), 
we may calculate the corresponding ``theoretical" branching ratios. These SM predictions require information on the CKM 
matrix elements $|V_{cb}|$ and $|V_{ub}|$ \cite{Ricciardi:2021shl}, 
as well as on the relevant non-perturbative hadronic form factors. In view of this feature, 
it is advantageous to combine these branching ratios with information from semileptonic $B$ decays.

The differential rate of a semileptonic $B\to P  \ell \bar{\nu}_{\ell}$ decay, where $P$ denotes a pseudoscalar meson,
can be written in the following form (neglecting lepton masses) \cite{Bigi:2016mdz, Li:2008tk, Colangelo:2006vm, Wang:2012ab}:
 \begin{equation}
\frac{\mathrm{d} \Gamma (\bar B \rightarrow P \ell \bar{\nu}_{\ell})}{\mathrm{d}q^2} = \frac{G^2_{\rm F} \left|V_{rb}\right|^2}{192 \, \pi^3} \, \left [ m_B \,
\Phi\left(\frac{m_P}{m_B},\frac{\sqrt{q^2}}{m_B}\right) \right]^3 \left[F^{B \rightarrow P}_1(q^2) \right]^2.
\label{eq:dg}
\end{equation}
Here the label $r=c,u$ distinguishes between $b \to c \ell \bar{\nu}_{\ell}$ and $b \to u\ell \bar{\nu}_{\ell}$ quark-level transitions, 
the phase-space function $\Phi(x,y)$ was introduced in Eq.\ (\ref{Phi-def}), and $F^{B \rightarrow P}_1(q^2)$ is the second form factor parametrising the corresponding quark-current matrix element (see Eq.~(\ref{FF-para})), satisfying the normalisation condition
\begin{equation}\label{FF-norm}
F^{B \rightarrow P}_1(0)=F^{B \rightarrow P}_0(0).
\end{equation}

In Eq.~(\ref{eq:dg}), we have again assumed the SM for the semileptonic decay amplitude. The corresponding modes may be affected by physics from beyond the SM \cite{Hiller:2018ijj,Fajfer:2019rjq,Bernlochner:2021vlv,Albrecht:2021tul}. However, it is possible to include NP effects in such decays \cite{Jung:2018lfu,Iguro:2020cpg,Fleischer:2021yjo,Banelli:2018fnx}. 
Should physics beyond the SM enter exclusively through couplings to heavy tau leptons, which is a popular scenario in 
the literature, the semileptonic decay rates into muons and electrons would still take the form in Eq.~(\ref{eq:dg}). When using experimental data, we shall only employ 
semileptonic $B_{(s)}$ decays into the light leptons $\ell=e,\mu$ (and hence neglected the lepton masses in Eq.~(\ref{eq:dg})).

\boldmath
\subsubsection{System of the $\bar{B^0_s} \rightarrow D_s^+ K^-$ and $\bar B^0_s \rightarrow D_s^+ \ell \bar{\nu}_{\ell}$ Decays}
\unboldmath
Let us first have a look at the $\bar{B^0_s} \rightarrow D_s^+ K^-$ channel, which originates from a $b \to c$ transition and is complemented through the semileptonic decay $\bar B^0_s \rightarrow D_s^+ \ell \bar{\nu}_{\ell}$. It is very useful to 
introduce a ratio of the following kind \cite{FST-BR, Neubert:1997uc, Beneke:2000ry}:
\begin{equation}
  R_{D_s^{+}K^{-}}\equiv\frac{\mathcal{B}(\bar{B}^0_s \rightarrow D_s^{+}K^{-})_{\rm th}}{{\mathrm{d}\mathcal{B}\left(\bar{B}^0_s \rightarrow D_s^{+}\ell^{-} \bar{\nu}_{\ell} \right)/{\mathrm{d}q^2}}|_{q^2=m_{K}^2}} ,
    \label{semi}
\end{equation}
where the differential branching ratio
\begin{equation}\label{Dg-Br-conv}
\frac{\mathrm{d}\mathcal{B}\left(\bar{B}^0_s \rightarrow D_s^{+}\ell^{-} \bar{\nu}_{\ell} \right)}{\mathrm{d}q^2} =\  \tau_{B_s} 
\left[\frac{\mathrm{d} \Gamma \left(\bar{B}^0_s \rightarrow D_s^{+}\ell^{-} \bar{\nu}_{\ell} \right)}{\mathrm{d}q^2}\right]
\end{equation}
is related to the differential rate through the $B_s$ average lifetime $\tau_{B_s}$. 
It should be noted that for $q^2=m_K^2$, the same phase space-functions enter the semileptonic and non-leptonic 
$\bar{B}^0_s$ decays, and that the CKM matrix element $|V_{cb}|$ cancels in the $R_{D_s^{+}K^{-}}$ ratio. 
Using Eqs.\ (\ref{BR-theo-single})--(\ref{Phi-def}) with (\ref{eq:dg}) and (\ref{Dg-Br-conv}), we obtain
\begin{equation}\label{RDsKm-expr}
R_{D_s^{+}K^{-}}=6 \pi^2 f_{K}^2 |V_{us}|^2 |a_{\rm 1 \, eff }^{D_s K}|^2  X_{D_s K},
\end{equation}
where 
\begin{equation}  \label{RSM}
    X_{D_s K}=\frac{(m_{B_s}^2 - m_{D_s}^2 )^2}{[m_{B_s}^2-(m_{D_s}+m_K)^2][m_{B_s}^2-(m_{D_s}-m_K)^2]}  \left[ \frac{F_0^{B_s \rightarrow D_s}(m_K^2)}{F_1^{B_s \rightarrow D_s}(m_K^2)} \right]^2.
\end{equation}
The product of the kaon decay constant
and the CKM factor $|V_{us}|$ can be extracted from data for leptonic $K$ decays, yielding
$f_K |V_{us}| = (35.09 \pm 0.04 \pm 0.04 ) \, {\rm MeV}$ \cite{Rosner:2015wva}. In Table~\ref{masses}, we collect the relevant meson masses. For the momentum transfer $q^2=m_K^2$, the 
ratio of hadronic form factors is close to the normalisation given in Eq.~(\ref{FF-norm}). Using the form-factor information from
lattice QCD studies \cite{Monahan:2017uby, McLean:2019qcx, Aoki:2019cca}, we obtain 
\begin{equation}
\left[ 
\frac{F_0^{B_s \rightarrow D_s}(m_K^2)}{F_1^{B_s \rightarrow D_s}(m_K^2)}
\right]
= 1.00 \pm 0.03.
\end{equation}

The differential rate of the $\bar{B}^0_s \rightarrow D_s^{+}\ell^{-} \bar{\nu}_{\ell}$ decay has recently 
been measured by the LHCb collaboration 
\cite{LHCb:2020cyw}. Applying the Caprini--Lellouch--Neubert (CLN) parametrisation \cite{Caprini:1997mu} of the relevant 
form factor with the parameters  
\begin{equation}
|V_{cb}|  = (41.4 \pm 1.3)\times 10^{-3},  \ \ \  G(0)=1.102 \pm 0.034, \ \ \ \rho^2=1.27\pm 0.05,
\end{equation} 
which result from the LHCb analysis \cite{LHCb:2020cyw}, we obtain 
\begin{equation} \label{eq:new}
\left.\frac{\mathrm{d}\mathcal{B}\left(\bar{B}^0_s \rightarrow D_s^{+}\ell^{-} \bar{\nu}_{\ell} \right)}{\mathrm{d}q^2}\right|_{q^2
=m_{K}^2} = 
\left(3.97 \pm 0.47 \right) \times 10^{-3} {\text{\ GeV}^{-2}} ,
 \end{equation}
 where we have used Eq.\ (\ref{Dg-Br-conv}) with the value of $\tau_{B_s}$ in Eq.\ (\ref{lifetimes}) to convert the differential rate into the differential  branching ratio. Here we have neglected correlations between the parameters to calculate the uncertainty. Taking them into account would reduce the error. However, for our numerical analysis, we prefer to use the larger uncorrelated error, also in view of the different form factors parametrizations that can be used. In the future, it would be really desirable if experimentalists measured the differential rates at the relevant $q^2$ bins. Then we would not have to use different form factor parametrisations. Combining this result with the theoretical branching ratio for the $\bar{B^0_s} \rightarrow D_s^+ K^-$ 
 mode in Table~\ref{tab:xi_BRth} yields
 \begin{equation}\label{RDsK-exp}
R_{D_s^{+}K^{-}}=0.05 \pm 0.01,
\end{equation}
 which allows us finally to determine  
 \begin{equation}\label{a-DsK-exp0}
|a_{\rm 1 \, eff }^{D_s K}| = 0.82 \pm 0.09.
\end{equation}
Using the expression in (\ref{a-eff-1-DsK}) with
\begin{equation}\label{rEdsK-val}
r_E^{D_sK}=1.00\pm0.08
\end{equation}
given in Eq.~(\ref{rE-values}) to take the contribution from the exchange topology into account, we obtain
 \begin{equation}\label{a-DsK-exp}
|a_{\rm 1}^{D_s K}| = 0.82 \pm 0.11.
\end{equation}
This result, which follows from the data and has a tiny dependence on hadronic form factors, has a surprisingly small central value
and differs from the theoretical expectation in Eq.\ (\ref{a1-1-pred}) at the $2.2\,\sigma$ level. We shall return to this puzzling feature
in Subsection~\ref{ssec:puzzles}.

\boldmath
\subsubsection{System of the $\bar{B^0_s} \rightarrow K^+ D_s^-$ and $\bar B^0_s \rightarrow K^+ \ell \bar{\nu}_{\ell}$ 
Decays}
\unboldmath
The decay $\bar{B^0_s} \rightarrow K^+ D_s^-$ originates from a $b \to u$ transition and is complemented through the 
semi-leptonic $\bar B^0_s \rightarrow K_s^+ \ell \bar{\nu}_{\ell}$ decay. In analogy to Eq.~(\ref{semi}), we introduce the ratio
\begin{equation}\label{R-2}
R_{K^{+}D_s^{-}}\equiv\frac{\mathcal{B}(\bar{B}^0_s \rightarrow D_s^{-}K^{+})_{\rm th}}{{\mathrm{d}\mathcal{B}}\left(\bar{B}^0_s \rightarrow K^{+}\ell^{-} \bar{\nu}_{\ell} \right)/{\mathrm{d}q^2}|_{q^2=m_{D_s}^2} },
\end{equation}
which takes the following form similar to Eq.~(\ref{RDsKm-expr}): 
\begin{equation}
R_{K^{+}D_s^{-}} = 6 \pi^2 f_{D_s}^2 |V_{cs}|^2 |a_{\rm 1  \, eff}^{K D_s}|^2 X_{K D_s},
\end{equation}
where
\begin{equation}
    X_{K D_s}=\frac{(m_{B_s}^2 - m_K^2)^2}{[m_{B_s}^2-(m_K+m_{D_s})^2][m_{B_s}^2-(m_K-m_{D_s})^2]} \left[ \frac{F_0^{B_s \rightarrow K}(m_{D_s}^2)}{F_1^{B_s \rightarrow K}(m_{D_s}^2)} \right]^2.
    \label{RSMK}
\end{equation}
The product of the $D_s$ decay constant and the CKM factor $|V_{cs}|$ can be determined from measurements of leptonic
$D_s$ decays, yielding $f_{D_s} |V_{cs}|=(250.9 \pm 4.0) \, {\rm MeV}$ \cite{Rosner:2015wva}. 

The semileptonic $\bar B^0_s \rightarrow K^+ \ell \bar{\nu}_{\ell}$ mode, which would also be very useful for an analysis of the 
$\bar{B}^0_s\to K^+K^-$ decay \cite{Fleischer:2016jbf}, has recently been observed by the LHCb collaboration with a first measurement of its branching ratio \cite{LHCb:2020ist}. However, the corresponding differential decay rate for various $q^2$ bins
was not reported. Consequently, we may not yet determine
the $R_{K^{+}D_s^{-}}$ ratio from the data. However, applying the $SU(3)$ flavour symmetry of strong interactions, we may replace
the semileptonic $B_s$ decay through its partner channel $B^0_d \rightarrow \pi^+ \ell^{-} {\bar{\nu}_{\ell}}$, for which we do have measurements of the differential rate by the BaBar and Belle collaborations  \cite{PDG,Amhis:2019ckw}. We introduce
\begin{equation}
R_{K^{+}D_s^{-}}^{{{SU(3)}}}\equiv\frac{\mathcal{B}(\bar{B}^0_s \rightarrow D_s^{-}K^{+})_{\rm th}}{{\mathrm{d}
\mathcal{B}}\left(\bar{B}^0 \rightarrow \pi^{+}\ell^{-} \bar{\nu}_{\ell} \right)/{\mathrm{d}
q^2}|_{q^2=m_{D_s}^2} } = 6 \pi^2 f_{D_s}^2 |V_{cs}|^2 |a_{\rm 1  \, eff}^{K D_s}|^2 X_{{{SU(3)}}},
\end{equation}
where
\begin{equation}
    X_{{{SU(3)}}}= \left(1 - \frac{m_K^2}{m_{B_s}^2} \right)^2   \frac{\left[ \Phi \left( \frac{m_{K}}{m_{B_s}}, \frac{m_{D_s}}{m_{B_s}} \right)\right]}{\left[\Phi \left(\frac{m_{\pi}}{m_{B}}, \frac{m_{D_s}}{m_{B_s}} \right)\right]^{3}}     \left[ \frac{F_0^{B_s \rightarrow K}(m_{D_s}^2)}{F_1^{B \rightarrow \pi}(m_{D_s}^2)}\right]^2.
    \label{RSMi}
\end{equation}
It should be noted that different phase-space factors enter in this expression, in contrast to the decay ratios considered above.
The ratio of form factors can be expressed as 
\begin{equation}\label{FF-rat-0-1}
\left[ \frac{F_0^{B_s \rightarrow K}(m_{D_s}^2)}{F_1^{B \rightarrow \pi}(m_{D_s}^2)} \right]^2 = \left[\frac{F_0^{B_s \rightarrow K}(m_{D_s}^2)}{F_1^{B_s \rightarrow K}(m_{D_s}^2)} \right]^2 \left[\frac{F_1^{B_s \rightarrow K}(m_{D_s}^2)}{F_1^{B \rightarrow \pi}(m_{D_s}^2)}\right]^2.
\end{equation}
The non-perturbative form factors have been determined with lattice QCD \cite{Flynn:2015mha, Bazavov:2019aom} and QCD 
light-cone sum rule analyses \cite{Khodjamirian:2017fxg, Imsong:2014oqa}. In view of the  currently large experimental
uncertainty of ${\mathcal{B}}(\bar{B}^0_s \rightarrow D_s^{-}K^{+})_{\rm th}$, we assume that the first ratio still satisfies the relation in 
Eq.~(\ref{FF-norm}) for $q^2=m_{D_s}^2$, i.e.\ is close to 1, which is actually in agreement with the analysis in 
Ref.~\cite{Flynn:2019jbg}. We have also applied this relation in Eq.\ (\ref{FF-SU3-0}). It would be important to have a dedicated lattice QCD study of this form-factor ratio in the future. The second form-factor ratio represents the $SU(3)$-breaking corrections. 
If we use the results given in Ref.~\cite{Khodjamirian:2017fxg} for $q^2=0$ and neglect again the evolution to $q^2=m_{D_s}^2$, 
as in the assumptions above, we have
\begin{equation}\label{eq:ff}
\left[ \frac{F_1^{B_s \rightarrow K}(m_{D_s}^2)}{F_1^{B \rightarrow \pi}(m_{D_s}^2)} \right] =1.12 \pm 0.12.
\end{equation}
For completeness, we perform a study of the $q^2$ evolution, applying the formalism given in Ref.~\cite{Khodjamirian:2017fxg}, and determine the form factors at $q^2=m_{D_s}^2$, finding the following values:
\begin{equation}
 {F_1^{B_s \rightarrow K}(m_{D_s}^2)}=0.366 \pm 0.028,  \qquad   {F_1^{B \rightarrow \pi}(m_{D_s}^2)}=0.323 \pm 0.028,
\end{equation}
leading to the ratio
\begin{equation}
\left[ \frac{F_1^{B_s \rightarrow K}(m_{D_s}^2)}{F_1^{B \rightarrow \pi}(m_{D_s}^2)} \right] =1.13 \pm 0.13,
\end{equation}
which is in excellent agreement with Eq.~(\ref{eq:ff}), showing that the effect of the $q^2$ evolution is negligible within the given errors.

Using the theoretical branching ratio for the $\bar{B^0_s} \rightarrow D_s^-K^+$ decay in Table~\ref{tab:xi_BRth} and the 
following experimental value of the differential semileptonic branching ratio \cite{Amhis:2019ckw}: 
\begin{equation}
{\mathrm{d}\mathcal{B}}\left(\bar{B}^0 \rightarrow \pi^{+}\ell^{-} \bar{\nu}_{\ell} \right)/{\mathrm{d}q^2}|_{q^2=m_{D_s}^2} = (7.14 \pm 0.46) \times 10^{-6} {\text{\ GeV}^{-2}},
\end{equation}
we obtain  
\begin{equation}\label{RKDs-exp}
R_{K^{+}D_s^{-}} = 3.64 \pm 1.70,
\end{equation}
yielding
\begin{equation}\label{a1KDs-0}
|a_{\rm 1  \, eff}^{K D_s}| =0.77 \pm 0.20
\end{equation}
with the help of the expressions given above. Using the range for $r_E^{KD_s}$ in Eq.~(\ref{rEKDs-range}) yields
\begin{equation}\label{a1KDs-extr}
|a_{\rm 1}^{K D_s}| =0.77 \pm 0.21.
\end{equation}
In comparison with Eq.~(\ref{a-DsK-exp}), the uncertainty is now significantly larger. However, we find again a similar pattern, with a 
central value smaller than the theoretical reference value in Eq.\ (\ref{a1-2-pred}). Although factorisation may not work as well as in the
$\bar{B^0_s} \rightarrow D_s^+K^-$ decay, this is yet another intriguing observation. 
It would be very important and interesting to reduce the corresponding uncertainties,
both the theoretical and the experimental ones, and to have a measurement of the differential 
$\bar B^0_s \rightarrow K_s^+ \ell \bar{\nu}_{\ell}$ decay rate available.

\subsection{Puzzling Patterns}\label{ssec:puzzles}
Let us now complement the results for the $|a_1|$ parameters of the $\bar{B^0_s} \rightarrow D_s^+K^-$ and 
$\bar{B^0_s} \rightarrow D_s^-K^+$ channels obtained in Section~\ref{ssec:SL} with the picture arising for decays with
similar dynamics. 
As we have seen in Section~\ref{ssec:fact}, the decay $\bar{B^0_d} \rightarrow D_d^+K^-$ originates from
$b\to c \bar{u} s$ processes in analogy to the $\bar{B^0_s} \rightarrow D_s^+K^-$ channel
but does not receive contributions from exchange topologies. Introducing 
  \begin{equation}\label{semi-DK}
  R_{D_d^{+}K^{-}}\equiv\frac{\mathcal{B}(\bar{B}^0_d \rightarrow D_d^{+}K^{-})}{{\mathrm{d}\mathcal{B}\left(\bar{B}^0_d \rightarrow D_d^{+}\ell^{-} \bar{\nu}_{\ell} \right)/{\mathrm{d}q^2}}|_{q^2=m_K^2}} 
  =6 \pi^2 f_{K}^2 |V_{us}|^2 |a_{\rm 1}^{D_d K}|^2  X_{D_d K}
 \end{equation}
 with
 \begin{equation}  \label{RSM-DK}
    X_{D_d K}=\frac{(m_{B_d}^2 - m_{D_d}^2 )^2}{[m_{B_d}^2-(m_{D_d}+m_K)^2][m_{B_d}^2-(m_{D_d}-m_K)^2]}  
    \left[ \frac{F_0^{B_d \rightarrow D_d}(m_K^2)}{F_1^{B_d \rightarrow D_d}(m_K^2)} \right]^2,
\end{equation}
which are the counterparts of Eqs.\ (\ref{RDsKm-expr}) and (\ref{RSM}), we may extract $|a_{\rm 1}^{D_d K}|$ from the data. 
Applying the CLN parametrisation  \cite{Caprini:1997mu} with the following parameters  \cite{Amhis:2019ckw}:
\begin{equation}
\eta_{\rm EW} G(1) |V_{cb}| = (42.00 \pm 1.00) \times 10^{-3}, \ \ \ \ \ \ \rho^2=1.131\pm 0.033,
\end{equation} 
and using the lifetime $\tau_{B_d}$ in Eq.~(\ref{lifetimes}), 
we get the following result for the differential branching ratio at the relevant value of $q^2=m_K^2$:
\begin{equation}
{{\mathrm{d}\mathcal{B}\left(\bar{B}^0_d \rightarrow D_d^{+}\ell^{-} \bar{\nu}_{\ell} \right)/{\mathrm{d}q^2}}|_{q^2=m_K^2}}=(3.65 \pm 0.23) \times 10^{-3} \ \text{GeV}^{-2} .
\end{equation}
Taking correlations between the parameters into account, we find a smaller uncertainty. However, as in Eq.\ (\ref{eq:new}), we prefer to keep the more conservative uncorrelated uncertainty. 
Using the form-factor ratio 
\begin{equation}
 \left[ \frac{F_0^{B_d \rightarrow D_d}(m_K^2)}{F_1^{B_d \rightarrow D_d}(m_K^2)} \right]=1
\end{equation}
in accordance with the normalisation condition (\ref{FF-norm}), we obtain
\begin{equation}
|a_{\rm 1}^{D_d K}|=0.83\pm0.05,
\end{equation}
which should be compared with the corresponding theoretical value in Eq.\ (\ref{a1-pred0}). We observe that the experimental 
central value is again significantly smaller, and encounter a discrepancy at the $4.8\, \sigma$ level.

For the $U$-spin partner $\bar{B}^0_d\to D_d^+\pi^-$ of the $\bar{B^0_s} \rightarrow D_s^+K^-$ channel  \cite{RF-BsDsK}, 
we introduce
  \begin{equation}\label{semi-Dpi}
  R_{D^{+}\pi^{-}}\equiv\frac{\mathcal{B}(\bar{B}^0_d \rightarrow D_d^{+}\pi^{-})}{{\mathrm{d}\mathcal{B}\left(\bar{B}^0_d \rightarrow D_d^{+}\ell^{-} \bar{\nu}_{\ell} \right)/{\mathrm{d}q^2}}|_{q^2=m_{\pi}^2}} 
  =6 \pi^2 f_{\pi}^2 |V_{ud}|^2 |a_{\rm 1 \, eff }^{D_d \pi}|^2  X_{D_d\pi}
 \end{equation}
 with
 \begin{equation}  \label{RSM-Dpi}
    X_{D_d \pi}=\frac{(m_{B_d}^2 - m_{D_d}^2 )^2}{[m_{B_d}^2-(m_{D_d}+m_{\pi})^2][m_{B_d}^2-(m_{D_d}-m_{\pi})^2]}  
    \left[ \frac{F_0^{B_d \rightarrow D_d}(m_{\pi}^2)}{F_1^{B_d \rightarrow D_d}(m_{\pi}^2)} \right]^2,
\end{equation}
where
\begin{equation}\label{a-eff-1-Dpi}
a_{\rm 1 \, eff }^{D_d \pi}=a_{1}^{D_d \pi} \left(1+\frac{E_{D_d \pi}}{T_{D_d \pi}}\right)
\end{equation}
takes also the exchange topology into account. Using $f_{\pi} |V_{ud}|=(127.13 \pm 0.02) \text{ MeV}$ \cite{Rosner:2015wva} and the experimental differential semileptonic branching ratio for $q^2=m_{\pi}^2$ \cite{Amhis:2019ckw}, 
\begin{equation}
{\mathrm{d}\mathcal{B}\left(\bar{B}^0_d \rightarrow D_d^{+}\ell^{-} \bar{\nu}_{\ell} \right)/{\mathrm{d}q^2}}|_{q^2=m_{\pi}^2} = \left(3.80 \pm 0.24 \right) \times 10^{-3} \ \text{GeV}^{-2},
\end{equation}
we find
\begin{equation}
|a_{\rm 1 \, eff }^{D_d \pi}|=0.83\pm0.03 \,.
\end{equation}
 If we assume $r_E^{D_d\pi}=r_E^{D_sK}$ with the numerical value in Eq.\ (\ref{rEdsK-val}), we get
 \begin{equation}
 |a_1^{D_d \pi}|=0.83\pm 0.07 \,,
 \end{equation}
which should be compared with the theoretical prediction in  Eq.\ (\ref{a1-pred0}). We observe again that the experimental value 
is much smaller, with a discrepancy at the $3.3\,\sigma$ level.
 
 The $\bar{B}^0_s\to D_s^+\pi^-$ decay differs from $\bar{B}^0_d\to D_d^+\pi^-$  
 only through the spectator quarks, and does not receive contributions from exchange topologies, thereby 
 representing a cleaner setting. In analogy to the discussion above, we introduce the ratio
 \begin{equation}   \label{semi-BsDspi}
  R_{D_s^{+}\pi^{-}}\equiv\frac{\mathcal{B}(\bar{B}^0_s \rightarrow D_s^{+}\pi^{-})_{\rm th}}{{\mathrm{d}
  \mathcal{B}\left(\bar{B}^0_s \rightarrow D_s^{+}\ell^{-} \bar{\nu}_{\ell} \right)/{\mathrm{d}q^2}}|_{q^2=m_{\pi}^2}} 
     =6 \pi^2 f_{\pi}^2 |V_{ud}|^2 |a_{\rm 1}^{D_s \pi}|^2  X_{D_s \pi}
\end{equation}
with
\begin{equation}  \label{RSM-Dspi}
    X_{D_s \pi}=\frac{(m_{B_s}^2 - m_{D_s}^2 )^2}{[m_{B_s}^2-(m_{D_s}+m_\pi)^2][m_{B_s}^2-(m_{D_s}-m_\pi)^2]}  \left[ \frac{F_0^{B_s \rightarrow D_s}(m_{\pi}^2)}{F_1^{B_s \rightarrow D_s}(m_{\pi}^2)} \right]^2.
\end{equation}
Using Eqs.\ (\ref{Dspi_th}) and (\ref{BR-values}), we find 
$\mathcal{B}(\bar{B}^0_s \rightarrow D_s^{+}\pi^{-})_{\rm th}=(2.99\pm0.23)\times10^{-3}$. Concerning the differential rate of the semileptonic $\bar{B}^0_s \rightarrow D_s^{+}\ell^{-} \bar{\nu}_{\ell} $ decay, 
we apply again the parameters of the CLN parametrisation  given by the LHCb collaboration 
in Ref.\ \cite{LHCb:2020cyw} with the $\bar B^0_s$ lifetime in Eq.\ (\ref{lifetimes}), yielding
\begin{equation}
{\mathrm{d}\mathcal{B}\left(\bar{B}^0_s \rightarrow D_s^{+}\ell^{-} \bar{\nu}_{\ell} \right)/{\mathrm{d}q^2}}|_{q^2=m_{\pi}^2} = \left(4.12 \pm 0.46 \right) \times 10^{-3} \ \text{GeV}^{-2}.
\end{equation}
Finally, we extract the following result from the data:
 \begin{equation}
 |a_1^{D_s \pi}|=0.87\pm0.06 \,.
 \end{equation}
Comparing this value with the theoretical prediction in Eq.\ (\ref{a1-pred0}), we observe that it is again too small, differing at the
$3.2\,\sigma$ level.

Finally, we consider the $\bar{B}^0_d\to \pi^+D_s^-$ decay, which is the counterpart of $\bar{B}^0_s\to K^+D_s^-$ and 
differs only through the spectator quarks. In particular, this mode does not have an exchange contribution. Introducing
\begin{equation}\label{RpiDs}
R_{\pi^{+}D_s^{-}}\equiv\frac{\mathcal{B}(\bar{B}^0_d \rightarrow \pi^{+}D_s^{-})}{{\mathrm{d}
\mathcal{B}}\left(\bar{B}^0 \rightarrow \pi^{+}\ell^{-} \bar{\nu}_{\ell} \right)/{\mathrm{d}
q^2}|_{q^2=m_{D_s}^2} } = 6 \pi^2 f_{D_s}^2 |V_{cs}|^2 |a_{\rm 1}^{\pi D_s}|^2 X_{\pi D_s}
\end{equation}
with
\begin{equation}\label{XpiDs}
    X_{\pi D_s}= \frac{(m_{B_d}^2 - m_{\pi}^2)^2}{[m_{B_d}^2-(m_{\pi}+m_{D_s})^2][m_{B_d}^2-(m_{\pi}-m_{D_s})^2]} \left[ \frac{F_0^{B_d \rightarrow \pi}(m_{D_s}^2)}{F_1^{B_d \rightarrow \pi}(m_{D_s}^2)} \right]^2,
\end{equation}
we find
\begin{equation}
|a_{\rm 1}^{\pi D_s}| = 0.78\pm0.05 \,,
\end{equation}
where we have used the experimental differential branching ratio \cite{Amhis:2019ckw}:
\begin{equation}
{\mathrm{d}
\mathcal{B}}\left(\bar{B}^0 \rightarrow \pi^{+}\ell^{-} \bar{\nu}_{\ell} \right)/{\mathrm{d}
q^2}|_{q^2=m_{D_s}^2} = \left(7.14 \pm 0.46 \right) \times 10^{-6} \ \text{GeV}^{-2}.
\end{equation}
As in Eq.~(\ref{a1KDs-0}), we have assumed that the form-factor ratio in Eq.~(\ref{XpiDs}) still satisfies the relation in 
Eq.~(\ref{FF-norm}) for $q^2=m_{D_s}^2$, i.e.\ is close to 1. It would be important to have a dedicated lattice QCD study 
of this form-factor ratio. In comparison with Eq.~(\ref{a1KDs-extr}), we have a consistent result although with significantly smaller uncertainty. The theoretical reference value in Eq.\ (\ref{a-1-rel-u}) differs from the experimental result at the $2.9\,\sigma$ level.

\begin{figure}[t]
	\centering
	\includegraphics[width = 0.57\linewidth]{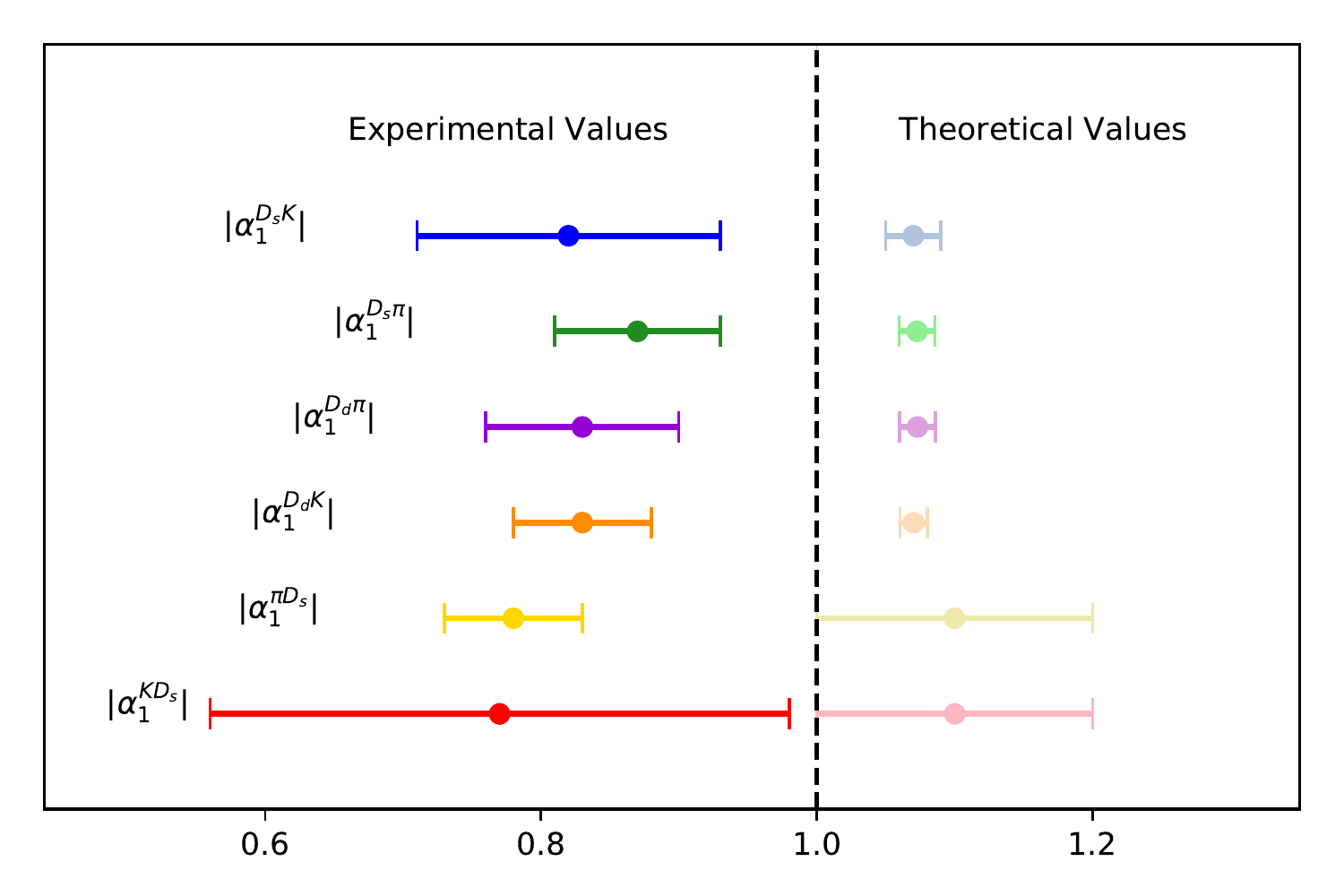}
	\caption{Experimental and theoretical SM values of the $|a_1|$ parameters for various decay processes as discussed in the text.} \label{fig:aval}
\end{figure}

In Fig.~\ref{fig:aval}, we show our results for the various $|a_1|$ parameters determined from the experimental data and compare them with the theoretical SM expectations. A similar pattern with values of $|a_1|$ smaller than one was found for $\bar B^0_d\to D_d^+\pi^-$ and  $\bar B^0_d\to D_d^+K^-$ decays in Ref.~\cite{FST-BR}, and has recently been identified and discussed in more detail within QCD factorisation also for $\bar B^0_s\to D_s^+\pi^-$ in Ref.~\cite{Bordone:2020gao}. This feature has led to recent analyses within scenarios for physics beyond the SM \cite{Iguro:2020ndk,Cai:2021mlt,Bordone:2021cca}. Within the SM, universal power-suppressed corrections of order $1/m_b$ could lead to a suppression of the $|a_1|$ parameters \cite{Huber:2016xod}. However, such effects would not allow us to accommodate the CP-violating observables of the $B^0_s\to D^\mp K^\pm$ system. 

The picture of the $|a_1|$ parameters of the $b \to c$ transitions is very intriguing, showing puzzling patterns with respect to the 
SM predictions of QCD factorisation. These decays are prime examples where this framework is expected to work
very well. Even in the $\bar B^0_s\to K^+D_s^{-}$ and $\bar B^0_d\to \pi^+D_s^{-}$ channels, where factorisation is on less solid ground, we find indications of a similar pattern (see also Ref.~\cite{DeBFKMST}). Interestingly, in our analysis in 
Section~\ref{ssec:fact}, we did not find any indication for an anomalous enhancement of the exchange topologies, 
which could -- in principle -- arise from large non-factorisable effects. A picture in favour of factorisation is also supported by the small 
strong phase $\delta_s$ in Eq.\ (\ref{LHCb-par-res}). These observations disfavour also anomalously enhanced power corrections. 

These observations are particularly exciting in view of the puzzling result for $\gamma$ which follows 
from the analysis of the CP-violating 
observables of the $\bar B^0_s\to D_s^\mp K^\pm$ decays in Section~\ref{sec:CPV}. Could they have a similar origin, arising from 
physics beyond the SM? Let us now generalise our discussion of the  $\bar B^0_s\to D_s^\mp K^\pm$ system to 
include NP effects.

\section{In Pursuit of New Physics}\label{sec:NP}

In view of these puzzles, we extend our analysis to include NP effects. As we have already noted at the end of Section 2, the anomalous patterns from the CP asymmetries would have to come from new CP-violating contributions at the decay amplitude level. Making our analysis for the branching ratios, we have actually found further puzzling patterns in the parameters $a_1$, which would be expected when allowing for new CP-violating contributions to the decay amplitudes, unless they enter in a very contrived way. Of course, the $\gamma$ puzzle could not be explained by large non-factorizable $\Lambda_{\text{QCD}}/m_b $ corrections, while - in principle - the $a_1$ values could be explained through such effects. The puzzles in the $a_1$ parameter by themselves would not require CP-violating effects while the intriguing result for $\gamma$ would necessarily require new CP-violating effects.

\subsection{New Physics Amplitudes}
The starting point of our analysis of NP effects is to generalise the transition amplitudes. Let us first have a look at the $\bar{B}^0_s$ and $B^0_s$ decays into the final state $D_s^{+} K^-$. We may write the corresponding decay amplitudes as follows:
\begin{equation}
A(\bar{B}^0_s \rightarrow D_s^+ K^-) = A(\bar{B}^0_s \rightarrow D_s^+ K^-)_{{\text{SM}}} \left[ 1 + \bar{\rho} \, e^{i \bar{\delta}} 
e^{+i \bar{\varphi}} \right]
\end{equation}
\begin{equation}
A({B}^0_s \rightarrow D_s^+ K^-) = A({B}^0_s \rightarrow D_s^+ K^-)_{{\text{SM}}} \left[ 1 + {\rho} \, e^{i {\delta}} e^{-i {\varphi}} \right] .
\end{equation}
Here $\bar\rho$ and $\rho$ describe the strength of the NP contributions to $b\to c \bar u s$ and $\bar b \to \bar u c \bar s$ quark-level transitions with respect to the corresponding SM amplitudes, respectively, with $\bar\delta$, $\delta$ denoting 
CP-conserving strong phases while $\bar\varphi$, $\varphi$ are CP-violating NP phases:
\begin{equation}
\bar{\rho} \, e^{i \bar{\delta}} e^{i \bar{\varphi}}  \equiv \frac{ A(\bar{B}^0_s \rightarrow D_s^+ K^-)_{{\text{NP}}} }{  A(\bar{B}^0_s \rightarrow D_s^+ K^-)_{{\text{SM}}} },
\quad
{\rho} \, e^{i {\delta}} e^{-i {\varphi}}\equiv \frac{A({B}^0_s \rightarrow D_s^+ K^-)_{{\text{NP}}} }{A({B}^0_s \rightarrow D_s^+ K^-)_{{\text{SM}}}}.
\end{equation}
These parameterisations of NP effects are actually more general than they may look at first sight, applying to situations where various NP contributions to the given quark-level transitions have the same CP-conserving or CP-violating phases. The point is that a general NP contribution
\begin{equation}
\sum_k \bar{\rho}_k \, e^{i \bar{\delta}_k} 
e^{i \bar{\varphi}_k}
\end{equation}
takes the forms
\begin{equation}
\sum_k \bar{\rho}_k \, e^{i \bar{\delta}_k} 
e^{+i \bar{\varphi}_k}=e^{i \bar{\delta}}  \sum_k \bar{\rho}_k \, 
e^{+i \bar{\varphi}_k} \equiv e^{i \bar{\delta}}\bar{\rho}e^{i \bar{\varphi}}
\end{equation}
and 
\begin{equation}
\sum_k \bar{\rho}_k \, e^{i \bar{\delta}_k} 
e^{-i \bar{\varphi}_k}=e^{i \bar{\delta}}  \sum_k \bar{\rho}_k \, 
e^{-i \bar{\varphi}_k} \equiv e^{i \bar{\delta}}\bar{\rho}e^{-i \bar{\varphi}}
\end{equation}
if all CP-conserving strong phases $ \bar{\delta}_k$ are equal to a universal phase $ \bar{\delta}$. In particular, the same $\bar{\rho}$ enters for the CP-conjugate expression as the general requirements for direct CP violation are not satisfied. In analogy, a similar structure arises if all CP-violation phases $\bar{\varphi}_k$ take the same value \cite{Fleischer:2001cw, Botella:2005ks}. 

The amplitudes for the CP-conjugate decay processes take the  form
\begin{equation}\label{NP-1}
A({B}^0_s \rightarrow D_s^- K^+) = A({B}^0_s \rightarrow D_s^- K^+)_{{\text{SM}}} \left[ 1 + \bar{\rho} \, e^{i \bar{\delta}} 
e^{-i \bar{\varphi}} \right]
\end{equation}
\begin{equation}\label{NP-2}
A(\bar {B}^0_s \rightarrow D_s^- K^+) = A(\bar {B}^0_s \rightarrow D_s^- K^+)_{{\text{SM}}} \left[ 1 + {\rho} \, 
e^{i {\delta}} e^{+i {\varphi}} \right]
\end{equation}
with
\begin{equation}
| A(\bar{B}^0_s \rightarrow D_s^+ K^-)_{{\text{SM}}} |=|A({B}^0_s \rightarrow D_s^-K^+)_{{\text{SM}}}|
\end{equation}
\begin{equation}
|A({B}^0_s \rightarrow D_s^+ K^-)_{{\text{SM}}}|=|A(\bar {B}^0_s \rightarrow D_s^- K^+)_{{\text{SM}}}|,
\end{equation}
reflecting the absence of direct CP violation in these decays in the SM.  Introducing the CP asymmetries
\begin{equation}
\bar{{\cal A}}^{\rm dir}_{\rm CP}\equiv \frac{|A({B}^0_s \rightarrow D_s^- K^+)|^2-|A(\bar{B}^0_s \rightarrow D_s^+ K^-)|^2}{|A({B}^0_s \rightarrow D_s^- K^+)|^2+|A(\bar{B}^0_s \rightarrow D_s^+ K^-)|^2} 
\end{equation}
\begin{equation}
{\cal A}^{\rm dir}_{\rm CP}\equiv \frac{|A({B}^0_s \rightarrow D_s^+ K^-)|^2-|A(\bar{B}^0_s \rightarrow D_s^- K^+)|^2}{|A({B}^0_s \rightarrow D_s^+ K^-)|^2+|A(\bar{B}^0_s \rightarrow D_s^- K^+)|^2} \, ,
\end{equation}
we obtain
\begin{equation}\label{Adir-expr}
\bar{{\cal A}}^{\rm dir}_{\rm CP}=\frac{2 \, \bar\rho\sin\bar\delta\sin\bar\varphi}{1+2 \, \bar\rho\cos\bar\delta\cos\bar\varphi + \bar\rho^2},
\quad
{\cal A}^{\rm dir}_{\rm CP}=\frac{2\,\rho\sin\delta\sin\varphi}{1+2 \, \rho\cos\delta\cos\varphi + \rho^2}.
\end{equation}
Consequently, we observe that NP may generate non-vanishing direct CP asymmetries, provided we have non-vanishing CP-conserving and CP-violating phases. 

Concerning the ratios $ R_{D_s^{+}K^{-}}$ and $R_{K^{+}D_s^{-}}$ introduced in Eqs.~(\ref{semi}) and (\ref{R-2}), respectively,
it is useful to generalize them through the following CP-averaged quantities:
\begin{equation}
 \langle R_{D_s K} \rangle \equiv\frac{\mathcal{B}(\bar{B}^0_s \rightarrow D_s^{+}K^{-})_{\rm th} + 
 \mathcal{B}({B}^0_s \rightarrow D_s^{-}K^{+})_{\rm th}}{\left[{\mathrm{d}\mathcal{B}\left(\bar{B}^0_s \rightarrow D_s^{+}\ell^{-} \bar{\nu}_{\ell} \right)/{\mathrm{d}q^2}}+ {\mathrm{d}\mathcal{B}\left({B}^0_s \rightarrow D_s^{-}\ell^{+} {\nu}_{\ell} \right)/{\mathrm{d}q^2}}\right]|_{q^2=m_{K}^2}} 
\end{equation}
\begin{equation}\label{R-2-av}
\langle R_{K D_s}\rangle
\equiv\frac{\mathcal{B}(\bar{B}^0_s \rightarrow K^{+}D_s^{-})_{\rm th}+
\mathcal{B}({B}^0_s \rightarrow K^{-}D_s^{+})_{\rm th}}{\left[{\mathrm{d}\mathcal{B}}\left(\bar{B}^0_s \rightarrow K^{+}\ell^{-} \bar{\nu}_{\ell} \right)/{\mathrm{d}q^2} + {\mathrm{d}\mathcal{B}}\left({B}^0_s \rightarrow K^{-}\ell^{+} {\nu}_{\ell} \right)/{\mathrm{d}q^2}\right]|_{q^2=m_{D_s}^2} }.
\end{equation}
In the case of vanishing direct CP violation in the corresponding modes, as in the SM considered in Subsection~\ref{ssec:SL}, 
we have $ \langle R_{D_s K } \rangle = R_{D_s^+ K^-}$ and
$\langle R_{K D_s }\rangle =R_{K^+D_s^-}$.  In the presence of NP contributions, it is useful to introduce the following quantities:
\begin{equation}\label{b-bar-def}
\bar b \equiv \frac{\langle R_{D_s K} \rangle}{6 \pi^2 f_{K}^2 |V_{us}|^2 |a_{\rm 1 \, eff}^{D_s K}|^2 X_{D_s K} }
=\frac{\langle\mathcal{B}(\bar {B}^0_s \rightarrow D_s^+K^-)_{\rm th}  \rangle}{\mathcal{B}(\bar{B}^0_s \rightarrow D_s^{+}K^{-})_{\rm th}^{\rm SM}}
=1+2 \, \bar\rho\cos\bar\delta\cos\bar\varphi + \bar\rho^2 
\end{equation}
\begin{equation}\label{b-def}
b\equiv \frac{\langle R_{K D_s}\rangle}{6 \pi^2 f_{D_s}^2 |V_{cs}|^2 |a_{\rm 1 \, eff}^{K D_s}|^2 X_{K D_s}}
=\frac{\langle\mathcal{B}(\bar{B}^0_s \rightarrow K^+D_s^-)_{\rm th}  \rangle}{\mathcal{B}(\bar{B}^0_s \rightarrow K^{+}D_s^{-})_{\rm th}^{\rm SM}}
= 1+2 \, \rho\cos\delta\cos\varphi + \rho^2 ,
\end{equation}
which allow us to probe the NP parameters, utilising the theoretical expectations of the parameters $|a_{1}^{D_s K}|$ and $|a_{1}^{K D_s}|$ with $r_E^{D_sK}$ and
$r_E^{K D_s}$, respectively, as input. 

Employing the direct CP asymmetry ${\cal A}^{\rm dir}_{\rm CP}$ 
and the branching ratio observable $b$, we may determine $\rho$ as function 
of the CP-violating phase $\varphi$ with the help of
\begin{equation}\label{rho-det}
\rho=\sqrt{u\pm\sqrt{u^2-w}},
\end{equation}
where
\begin{equation}
u\equiv b-1+2 \cos^2 \varphi 
\end{equation}
and
\begin{equation}
w\equiv\left(b-1\right)^2+\left(\frac{b \, {\cal A}^{\rm dir}_{\rm CP}}{\tan\varphi}\right).
\end{equation}
Similar expressions hold for the CP-conjugate quantities, allowing the extraction of the NP parameter 
$\bar{\rho}$ as function of $\bar{\varphi}$.

\boldmath
\subsection{Interference Effects Through $B^0_s$--$\bar B^0_s$ Mixing}
\unboldmath
For the CP-violating phenomena in the $B^0_s\to D_s^\mp K^\pm$ system, interference effects
between the different decay paths through $B^0_s$--$\bar B^0_s$ mixing play the essential role, as we have seen in 
Section~\ref{sec:CPV}. Let us now generalise these considerations to allow for NP contributions with new sources of CP 
violation. The key information is encoded in the observables $\xi$ and $\bar\xi$, which we may write using Eqs.~(\ref{xi-ME}) 
and (\ref{xi-bar-ME}) with (\ref{NP-1}) and (\ref{NP-2}) as follows:
\begin{equation}
\xi=- e^{-i(\phi_s + \gamma)} \left[ \frac{1}{x_s e^{i \delta_s}} \right]
\left[\frac{1 + \bar{\rho} \, e^{i \bar{\delta}} e^{+i \bar{\varphi}}}{1 + {\rho} \, e^{i {\delta}} e^{-i {\varphi}}}\right]
=-|\xi|e^{-i\delta_s}e^{-i(\phi_s+\gamma)}e^{i\Delta\varphi}
\end{equation}
\begin{equation}
\bar{\xi} =  - e^{-i(\phi_s + \gamma)} \left[{x_s e^{i \delta_s}} \right]
\left[\frac{1 + {\rho} \, e^{i {\delta}} e^{+i {\varphi}}}{1 + \bar{\rho} \, e^{i \bar{\delta}} 
e^{-i \bar{\varphi}}}\right]
=-|\bar{\xi}|e^{+i\delta_s}e^{-i(\phi_s+\gamma)}e^{i\Delta\bar{\varphi}},
\end{equation}
where
\begin{equation}\label{tan-varphi}
\tan\Delta\varphi=\frac{\rho\sin(\varphi-\delta)+\bar{\rho}\sin(\bar{\varphi}+\bar{\delta})+\bar{\rho}\rho\sin(\bar{\delta}-\delta+\bar{\varphi}+\varphi)}{1+\rho\cos(\varphi-\delta)+\bar{\rho}\cos(\bar{\varphi}+\bar{\delta})+\bar{\rho}\rho\cos(\bar{\delta}-\delta+\bar{\varphi}+\varphi)}
\end{equation}
and
\begin{equation}\label{tan-varphi-bar}
\tan\Delta\bar{\varphi}=\frac{\bar{\rho}\sin(\bar{\varphi}-\bar{\delta})+{\rho}\sin({\varphi}+{\delta})+{\rho}\bar{\rho}\sin({\delta}-\bar{\delta}+{\varphi}+\bar{\varphi})}{1+\bar{\rho}\cos(\bar{\varphi}-\bar{\delta})+{\rho}\cos({\varphi}+{\delta})+{\rho}\bar{\rho}\cos({\delta}-\bar{\delta}+{\varphi}+\bar{\varphi})}.
\end{equation}

Since the hadronic parameter $x_s$ with its CP-conserving strong phase $\delta_s$ cancels in the product of these observables, this 
combination is central for studying CP violation:
\begin{equation}
\xi \times \bar{\xi}  
= e^{-i2 (\phi_s + \gamma)}
\Biggl[\frac{1 + {\rho} \, e^{i {\delta}} e^{+i {\varphi}}}{1 + {\rho} \, e^{i {\delta}} e^{-i {\varphi}}}
\Biggr]
\Biggl[\frac{1 + \bar{\rho} \, e^{i \bar{\delta}} e^{+i \bar{\varphi}}}{1 + \bar{\rho} \, e^{i \bar{\delta}} 
e^{-i \bar{\varphi}}} \Biggr].
\end{equation}

Using the relation
\begin{equation}
\frac{1 + {\rho} \, e^{i {\delta}} e^{+i {\varphi}}}{1 + {\rho} \, e^{i {\delta}} 
e^{-i {\varphi}}} = e^{-i\Delta\Phi} \sqrt{\frac{1- {{\cal A}}^{\rm dir}_{\rm CP}}{1+{{\cal A}}^{\rm dir}_{\rm CP}}} 
\end{equation}
with
\begin{equation}
\cos\Delta\Phi= \sqrt{\frac{1- {{\cal A}}^{\rm dir}_{\rm CP}}{1+{{\cal A}}^{\rm dir}_{\rm CP}}} \left[ \frac{1+2\rho\cos\delta\cos\varphi+
\rho^2\cos2\varphi}{1+2\rho\cos(\delta-\varphi)+\rho^2} \right]
\end{equation}
and
\begin{equation}
\tan\Delta\Phi=-\left[\frac{2\rho\cos\delta\sin\varphi+
\rho^2\sin2\varphi}{1+2\rho\cos\delta\cos\varphi+
\rho^2\cos2\varphi}\right],
\end{equation}
as well as the corresponding counterparts for the phase $\Delta\bar{\Phi}$ that 
are related to $\bar\rho$, $\bar\varphi$ and $\bar{{\cal A}}^{\rm dir}_{\rm CP}$ in an analogous way, we obtain 
\begin{equation}\label{xi-prod0}
\xi \times \bar{\xi}  = e^{-i2 (\phi_s + \gamma)}\sqrt{\left[ \frac{1- {{\cal A}}^{\rm dir}_{\rm CP}}{1+{{\cal A}}^{\rm dir}_{\rm CP}}\right]
\left[\frac{1- {\bar{\cal A}}^{\rm dir}_{\rm CP}}{1+{\bar{\cal A}}^{\rm dir}_{\rm CP}}\right]}e^{-i(\Delta\Phi+\Delta\bar{\Phi})},
\end{equation}
where the CP-violating NP phase shifts satisfy the sum rule
\begin{equation}
\Delta\Phi+\Delta\bar{\Phi} = -\left(\Delta\varphi+\Delta\bar{\varphi}\right).
\end{equation}
Let us simplify the expression in Eq.~(\ref{xi-prod0}) 
further using the observable $C$ in Eq.~(\ref{obs-xi}) and its CP-conjugate. Writing 
\begin{equation}
\left|\xi\times\bar\xi\right|^2=
\left[ \frac{1- {{\cal A}}^{\rm dir}_{\rm CP}}{1+{{\cal A}}^{\rm dir}_{\rm CP}}\right]
\left[\frac{1- {\bar{\cal A}}^{\rm dir}_{\rm CP}}{1+{\bar{\cal A}}^{\rm dir}_{\rm CP}}\right]
=1+\epsilon,
\end{equation} 
we obtain
\begin{equation}
-\frac{1}{2}\, \epsilon = \frac{C+\bar{C}}{\left(1+C\right)\left(1+\bar{C}\right)} = 
{\cal A}^{\rm dir}_{\rm CP} + {\bar{\cal A}}^{\rm dir}_{\rm CP} + {\cal O}(({\cal A}^{\rm dir}_{\rm CP})^2),
\end{equation}
generalising the relation in Eq.~(\ref{C-rel}) which was assumed by the LHCb collaboration
in Ref.~\cite{LHCb-BsDsK}. Finally, we arrive at the following expression:
\begin{equation}
\xi \times \bar{\xi}  = \sqrt{1-2\left[\frac{C+\bar{C}}{\left(1+C\right)\left(1+\bar{C}\right)}
\right]}e^{-i\left[2 (\phi_s + \gamma)+\Delta\Phi+\Delta\bar{\Phi}\right]}.
\end{equation}
This is the generalisation of Eq.~(\ref{multxi}). The corresponding product of $\xi$ and $\bar{\xi}$ can still be 
determined through the observables of the time-dependent rate asymmetries of the $B^0_s\to D_s^\mp K^\pm$ system. 
We observe that possible corrections to the relations in Eqs.~(\ref{xi-rel}) and (\ref{C-rel}) are now manifestly included, and that the 
UT angle $\gamma$ actually enters with a shift due to the CP-violating NP phases, thereby resulting in the  ``effective" angle
\begin{equation}\label{gamma-eff}
\gamma_{\rm eff}\equiv \gamma+\frac{1}{2}\left(\Delta\Phi+\Delta\bar{\Phi}\right)=
\gamma-\frac{1}{2}\left(\Delta\varphi+\Delta\bar{\varphi}\right).
\end{equation}
In the future, it would be important to generalise the experimental analyses of the 
$\bar{B}^0_s \rightarrow D_s^+ K^-$ and ${B}^0_s \rightarrow D_s^+ K^-$ decays and their CP conjugates correspondingly.

\subsection{New Physics Analysis of the Data}
\subsubsection{Preliminaries}
Let us now apply the model-independent formalism developed above and come back to the intriguing patterns in the data that we 
encountered in Sections~\ref{sec:CPV} and \ref{sec:BR}, interpreting them in terms of NP contributions with new sources of 
CP violation. In order to have a scenario in agreement with the relation in Eq.\ (\ref{C-rel}),
which was assumed by LHCb in Ref.~\cite{LHCb-BsDsK}, we make the following assumption:
\begin{equation}\label{del-rel}
\delta=\bar{\delta}=0^\circ.
\end{equation}
It implies vanishing direct CP asymmetries $\bar{{\cal A}}^{\rm dir}_{\rm CP}$ and ${\cal A}^{\rm dir}_{\rm CP}$, as can be seen in 
Eq.~(\ref{Adir-expr}), which would be consistent with the measurements of such observables in tree-diagram-like decays of the 
kind $B\to D K$ within the current uncertainties \cite{PDG}. 

In such a scenario, the NP amplitude would enter with the same CP-conserving 
strong phase as the SM amplitude, which may actually well be the case. We will describe possible relative minus 
signs through the CP-violating phases $\varphi$ and $\bar{\varphi}$, i.e.\ through terms of $180^\circ$. The strong 
phases would be generated through non-factorisable effects arising in the hadronic matrix elements of four-quark operators. In view 
of the discussion of factorisation in Subsection~\ref{ssec:fact}, we would expect smallish phases for the decays at hand,
in particular for the $b\to c $ mode $\bar B^0_s\to D_s^+K^-$. The result for the strong phase difference $\delta_s$ 
in Eq.~(\ref{LHCb-par-res}) would be fully consistent with this picture within the uncertainties.

\subsubsection{New Physics Parameter Correlations}
\begin{figure} 
	\centering
	\includegraphics[width = 0.47\linewidth]{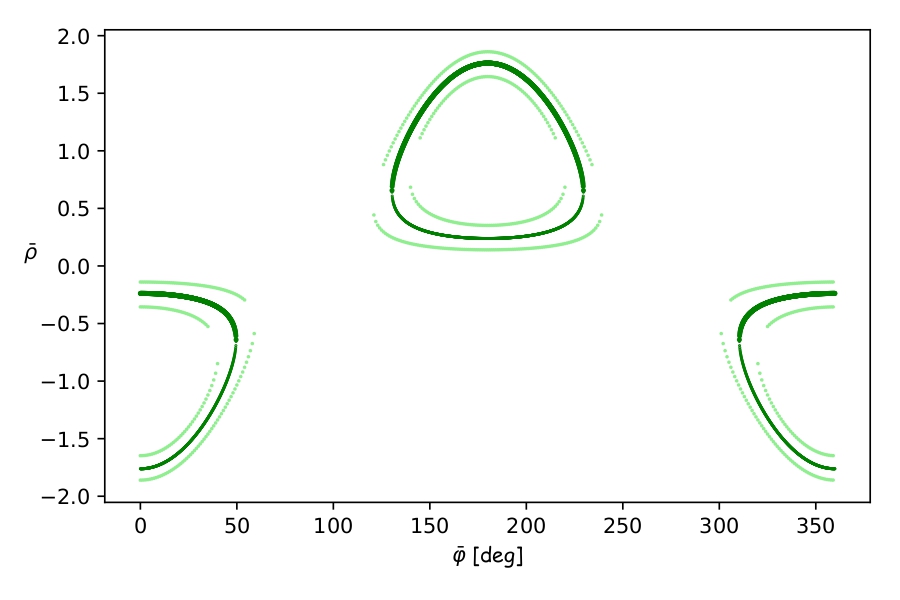}
	\includegraphics[width = 0.47\linewidth]{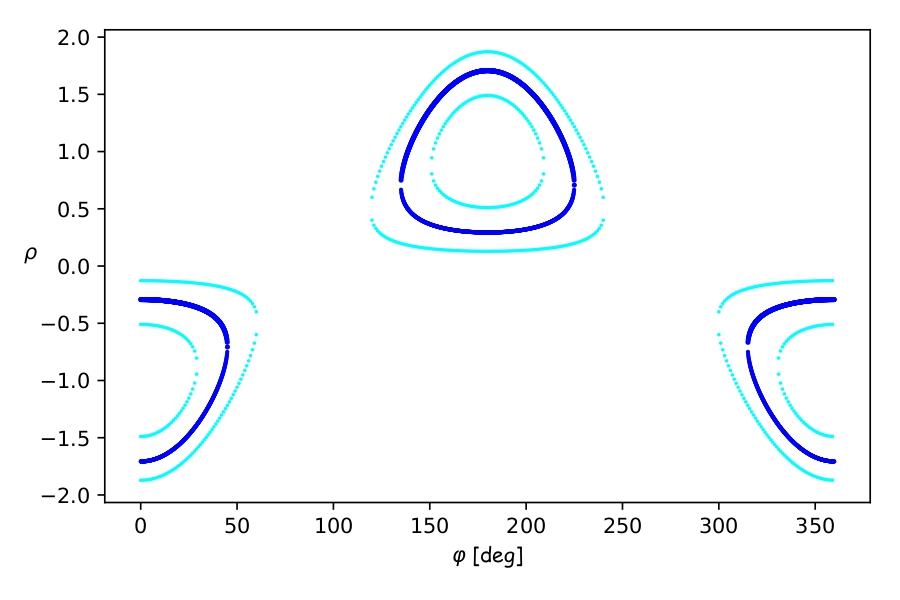}
	\caption{In the left panel, we show $\bar{\rho}$ as function of the CP-violating phase $\bar{\varphi}$, following from the branching ratio observable $\bar{b}$. In the right panel, we show the corresponding analysis for $\rho$ and $\varphi$, using $b$. } \label{fig:br-analysis}
\end{figure}

As the first observables, we consider the branching ratio information encoded in Eqs.~(\ref{b-bar-def}) and (\ref{b-def}). Assuming
the relation in Eq.~(\ref{del-rel}), we obtain the expressions 
\begin{eqnarray}
        \bar{\rho} &= & - \cos{\bar{\varphi}} \pm \sqrt{\bar{b} -\sin^2\bar{\varphi}}     \label{rhos} \\
      \rho & = &- \cos{\varphi} \pm \sqrt{b -\sin^2{\varphi}},    \label{rhoi}
\end{eqnarray}
which are in agreement with Eq.~(\ref{rho-det}) for vanishing direct CP asymmetries. Using the experimental values in
Eqs.\ (\ref{RDsK-exp}) and (\ref{RKDs-exp}) with the theoretical expectations of the $|a_1|$ parameters in 
Eqs.\ (\ref{a1-1-pred}) and (\ref{a1-2-pred}), respectively, complemented with the numerical values of $r_E^{KD_s}$ and $r_E^{D_sK}$ in (\ref{rEKDs-range}) and (\ref{rEdsK-val}), we obtain 
\begin{equation} \label{eq:a1th}
|a_{\rm 1 \, eff}^{D_s K}| = 1.07 \pm 0.09, \ \ \ \ \ |a_{\rm 1 \, eff}^{K D_s}|=1.1 \pm 0.13,
\end{equation}
yielding
\begin{equation}
\bar b=0.58 \pm 0.16,\ \ \ \ \ {b}=0.50 \pm 0.26.
\end{equation} 
Finally, applying Eqs.\ (\ref{rhos}) and (\ref{rhoi}), we get the constraints on the NP parameters shown in 
Fig.~\ref{fig:br-analysis}. Here the green (blue) contour shows the parameter $\rho$ $(\bar{\rho})$ as a function of the phase $\varphi$ $(\bar{\varphi})$ for the central value of the observable $b$ $(\bar{b})$. We include also the uncertainties, varying the values of the observable $b$ $(\bar{b})$ within the $1\,\sigma$ range, which leads to the contours in lighter colours.

Let us now use in addition the observables of the time-dependent $B^0_s\to D_s^\mp K^\pm$ decay rates. Looking
at Eqs.\ (\ref{tan-varphi}) and (\ref{tan-varphi-bar}) with (\ref{gamma-eff}), we observe that Eq.~(\ref{del-rel}) implies 
\begin{equation}
\Delta\varphi=\Delta\bar{\varphi} = \gamma-\gamma_{\rm eff}
\end{equation}
with
\begin{equation}\label{tan-varphi-rel}
\tan\Delta\varphi=\frac{\rho\sin\varphi+\bar{\rho}\sin\bar{\varphi}+\bar{\rho}\rho\sin(\bar{\varphi}+\varphi)}{1+\rho\cos\varphi +
\bar{\rho}\cos\bar{\varphi}+\bar{\rho}\rho\cos(\bar{\varphi}+\varphi)}.
\end{equation}
Using the result in Eq.\ (\ref{gamma-res-1}), which actually corresponds to the effective angle $\gamma_{\rm eff}$, and applying
$\gamma=(70 \pm 7)^\circ$, summarising the picture from analyses of CP violation in tree-level 
decays of the kind $B\to DK$ \cite{Amhis:2019ckw}, we obtain 
\begin{equation}\label{DelVarPhi}
\Delta \varphi = -(61 \pm 20)^\circ.
\end{equation}
It should be emphasised that this NP phase shift was extracted in a {\it theoretically clean} way from the data. 
In particular, it does not rely on SM predictions of the observables $\xi$ and $\bar{\xi}$, which is a very important feature
and non-trivial finding.

\begin{figure}[t]
		\centering
	\includegraphics[width = 0.55\textwidth]{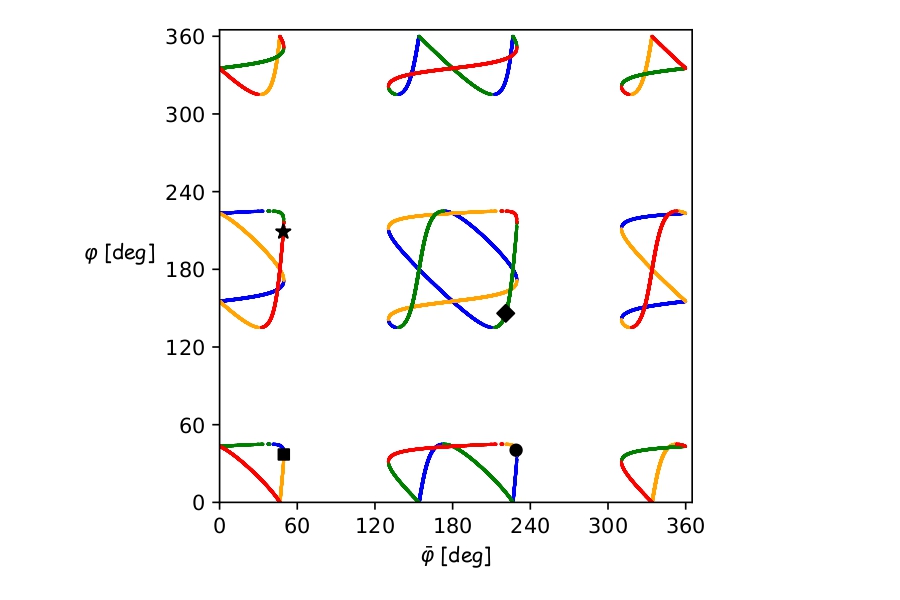} \hspace{-1.2cm}
	\hspace{-0.9cm}\includegraphics[width = 0.56\textwidth]{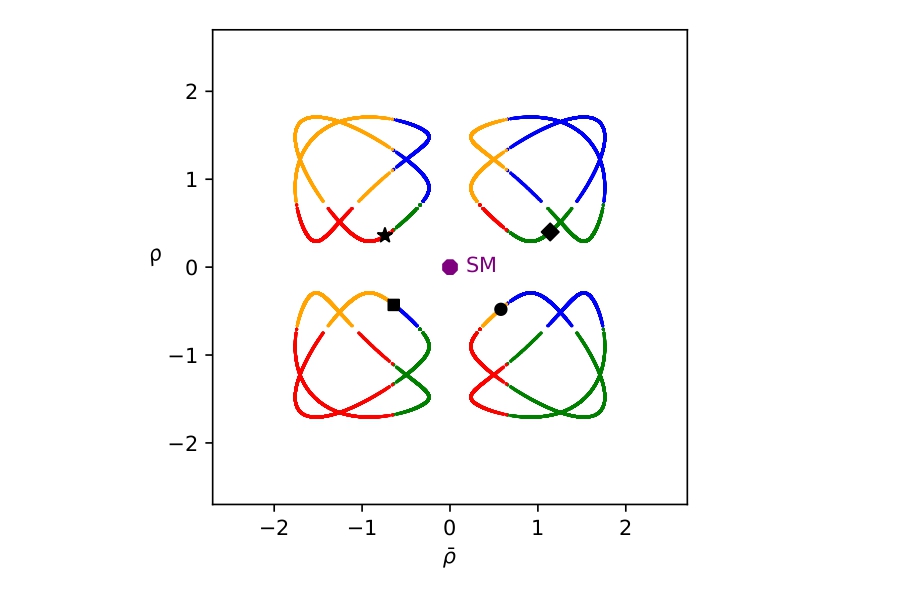} 
	\vspace*{-0.5truecm}
	\caption{Correlations in the $\bar{\varphi}$--$\varphi$ plane (left) and the $\bar{\rho}$--$\rho$ plane (right) 
	for the central values of the current data, as discussed in the text. We have indicated four illustrative points to show
	their appearance in the two correlations, and have made the SM point explicit.}\label{fig:centrrhophi}
\end{figure}

Employing again $\bar b$ and $b$ (see Eqs.~(\ref{rhos}) and (\ref{rhoi})), we may express $\bar{\rho}$ and ${\rho}$ in 
Eq.\ (\ref{tan-varphi-rel}) as functions of $\bar{\varphi}$ and $\varphi$, respectively, thereby allowing us to determine 
${\varphi}$ as a function of $\bar\varphi$ from the experimental value of $\Delta \varphi$ given in Eq.~(\ref{DelVarPhi}). With the help of 
Eqs.~(\ref{rhos}) and (\ref{rhoi}), we may then also determine the corresponding correlation in the $\bar{\rho}$--$\rho$ 
plane, where each point corresponds to a value of the CP-violating NP phase $\bar\varphi$.

In Fig.~\ref{fig:centrrhophi}, we show the corresponding
correlations for the central values of the current data. The contours in four different colours correspond to the four different combinations of the product $\rho \bar{\rho}$  in Eq.~(\ref{tan-varphi-rel}), arising from the two possible values that each one of $\rho$ and $\bar{\rho}$ can take due to the different sign before the square root in Eqs.\ (\ref{rhos}) and (\ref{rhoi}). 

As examples, we pick some values from the correlation in the $\bar\varphi$--$\varphi$ plane and show the corresponding values in 
the $\bar\rho$--$\rho$ plane. We illustrate these points in Fig.~\ref{fig:centrrhophi} as a square, circle, diamond and star, corresponding to the following NP parameter sets:
\begin{equation}\label{example1}
(\rho,\varphi)=(-0.43,37.0^\circ), \quad (\bar\rho,\bar\varphi)=(-0.64, 49.6^\circ) 
\end{equation}
\vspace*{-0.8truecm}
\begin{equation}
(\rho,\varphi)=(-0.48,40.3^\circ), \quad (\bar\rho,\bar\varphi)=(0.58, 229.0^\circ)
\end{equation}
\begin{equation}
(\rho,\varphi)=(0.40,146.0^\circ), \quad (\bar\rho,\bar\varphi)=(1.14, 221.0^\circ)
\end{equation}
\begin{equation}\label{example4}
\,\, (\rho,\varphi)=(0.36 ,209.0^\circ), \quad (\bar\rho,\bar\varphi)=(-0.74,49.2^\circ).
\end{equation}

\begin{figure}[t]
	\centering
\includegraphics[width = 0.65\linewidth]{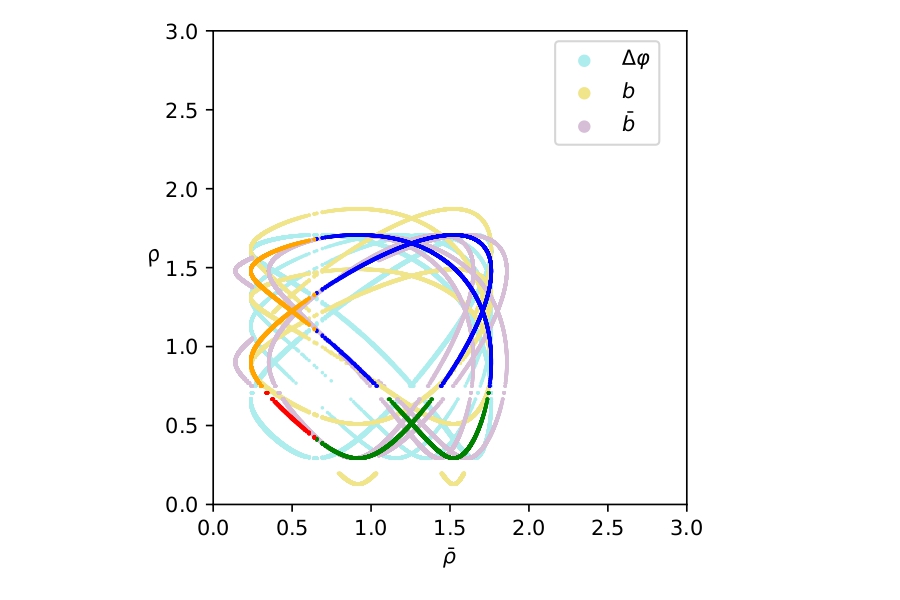}
\vspace*{-0.3truecm}
	\caption{The correlations in the $\bar{\rho}$--$\rho$ plane including the uncertainties of the input quantities $\bar{b}$, $b$ 
	and $\Delta \varphi$.}
	\label{fig:centrrhophi-1}
\end{figure}

\subsubsection{Discussion}
In the correlations in the $\bar{\varphi}$--$\varphi$ plane, interestingly at least one of the CP-violating phases has to take a non-trivial value, thereby illustrating the need for new sources of CP violation. In the $\bar{\rho}$--$\rho$ plane, the SM point corresponding to
the origin $(0,0)$ is excluded. 
The gaps between the various contours are in particular related to the values of $\bar b$ and $b$, which are smaller than one, and
arise from the algebraic structure of the underlying expressions. We notice that $\rho$, $\bar{\rho}$ are bounded to values below
two. Interestingly, values as small as in the regime around 0.5 could accommodate the central values of the current data, thereby resolving the puzzling patterns in the measurements of CP violation as well as in the branching ratios. Here we would then have NP contributions at the level of $50\%$ of the SM amplitudes. This feature is also nicely reflected by the NP parameter sets in 
Eqs.\ (\ref{example1})--(\ref{example4}).

In Fig.~\ref{fig:centrrhophi-1}, we illustrate the impact of the uncertainties of the input quantities $\bar{b}$, $b$  and $\Delta \varphi$
on the contours in the $\bar{\rho}$--$\rho$ plane. Focusing only on the positive values of $\rho$ and $\bar{\rho}$, we vary the values of each of the above three parameters separately and obtain the contours which are denoted with pale colours. Every different pale colour corresponds to one of the three parameters. We could now accommodate the data with NP contributions as small as about
$30\%$ of the SM amplitudes, which is an exciting observation. 

In Refs.~\cite{Iguro:2020ndk,Cai:2021mlt,Bordone:2021cca}, specific NP scenarios which may affect 
$B_{(s)}\to D_{(s)}\pi$ and $B\to DK$ decays were recently discussed in view of the puzzling patterns in the branching ratios of these
modes (see Section~\ref{sec:BR}). Such kind of physics beyond the SM, involving, for instance, left-handed $W'$ bosons \cite{Iguro:2020ndk}, would also enter the $B^0_s\to D_s^\mp K^\pm$ system, 
which offers an exciting probe for CP-violating NP phases as we have demonstrated in our analysis. We consider these first proposed models, which are facing challenges with high-$p_{\rm T}$ collider data for direct NP searches by the ATLAS and CMS collaborations \cite{Bordone:2021cca}, 
as interesting illustrations of possible scenarios. The model-independent NP analysis presented above will serve a benchmark in the future as the data improve, helping us to narrow down specific scenarios and models.

It is interesting to note that in studies to extract $\gamma$ from simultaneous fits to a variety of $B$ decays, such as in Ref.\ \cite{LHCb:2021dcr}, NP effects may average out to some extend, thereby resulting in an effective angle with NP contributions 
which -- in contrast to the situation in Eq.\ (\ref{gamma-eff}) -- cannot transparently be quantified. On the other hand, instead of such an involved fit, it would be crucial to search for patterns in conflict with the SM in the individual $\gamma$ determinations following from the various channels, aiming to perform them with highest precision. Our strategy for analysing CP violation in the $B^0_s\to D_s^\mp K^\pm$ system and the corresponding branching ratios is a prime example in this respect.

\section{Conclusions}\label{sec:concl}
Decays of $\bar{B}^0_s$ and $B^0_s$ mesons into the final state $D_s^+ K^-$ and its CP conjugate $D_s^- K^+$ provide an important laboratory for the testing the quark-flavour sector of the SM. Due to interference effects between these decay channels that are 
induced by $B^0_s$--$\bar B^0_s$ mixing, CP-violating asymmetries arise, which allow a theoretically clean determination of the UT angle
$\gamma$ within the SM. We have performed an analysis of LHCb measurements of these CP asymmetries, paying
special attention to the resolution of discrete ambiguities. We have resolved a final discrete ambiguity, leaving us with
$\gamma=\left(131^{+17}_{-22}\right)^\circ$, where we have used a recent result for $\phi_s$ taking penguin corrections into account.  
The large value of $\gamma$ is surprising in view of the range around $70^\circ$ following from tree 
decays of the kind $B\to DK$ and global analyses of the UT, differing at the $3\sigma$ level. It is important to emphasize that this tension could not 
be explained through non-factorizable effects.
 
Complementing the CP-violating observables with the measurement of an averaged branching ratio, we have extracted the 
individual branching ratios of the 
$\bar{B}^0_s\to D_s^+ K^-$ and $B^0_s\to D_s^+ K^-$ decays from the data, thereby disentangling the interference effects
between the different decay paths. In this analysis, we have also properly taken the effects of $B^0_s$--$\bar B^0_s$ mixing 
into account. 
 
In the experimental study of the $B^0_s\to D_s^\mp K^\pm$ system, the LHCb collaboration has assumed that direct 
CP violation vanishes in these channels, as would be the case in the SM. However, NP effects with new sources of CP violation 
could actually generate such CP asymmetries. It would be interesting to constrain them experimentally in the future. 
We have presented the corresponding generalised formalism in this paper. 
 
In order to minimise the impact of the uncertainties of hadronic form factors and the CKM matrix elements $|V_{cb}|$ and $|V_{ub}|$, 
we have introduced ratios of the 
 $\bar{B}^0_s\to D_s^+ K^-$ and $\bar{B}^0_s\to K^+D_s^-$  branching ratios with the differential rates of semileptonic
 $\bar{B}^0_s\to D_s^+ \ell^-\bar{\nu}_\ell$ and $\bar{B}^0_s\to K_s^+ \ell^-\bar{\nu}_\ell$ decays, respectively. These quantities
 allow clean determinations of the parameters $|a_1^{D_sK}|$ and $|a_1^{KD_s}|$, which characterise non-factorizable QCD
 effects. For the former $b\to c$ mode, we find  $|a_1^{D_sK}|= 0.82 \pm 0.11$. This value, which has a solid interpretation within QCD factorisation, 
 is intriguing as it differs at the $2.2 \,\sigma$ level from the corresponding theoretical expectation. In the case of $\bar{B}^0_s\to D_s^+ K^-$, factorisation is expected to work very well, similar to 
 $\bar B^0_d\to D_d^+\pi^-$,  $\bar B^0_s\to D_s^+\pi^-$ and  $\bar B^0_d\to D_d^+K^-$ channels. Interestingly, in the latter decays, 
a similar pattern arises, as was found in the previous literature; for $\bar{B^0_s} \rightarrow D_s^+K^-$, we obtain a discrepancy of $4.8\sigma$ with respect to the SM predictions utilizing QCD factorization. We confirm this picture in the $\bar{B}^0_s\to D_s^+ K^-$ channel, thereby complementing the puzzling result for $\gamma$ following from the CP asymmetries. In contrast to the picture of $\gamma$, unaccounted non-factorizable effects could accommodate the $|a_1|$ values. However, such corrections are not favoured by the measured value of the strong phase difference $\delta_s$. 
In the case of the $b\to u$ transition $\bar{B}^0_s\to K^+D_s^-$, the differential rate of the semileptonic partner $\bar{B}^0_s\to K^+ \ell^-\bar{\nu}_\ell$ has not yet been measured. However, replacing it
through the $SU(3)$-related channel $\bar{B}^0_d\to \pi^+ \ell^-\bar{\nu}_\ell$, we obtain $|a_{1}^{K D_s}|=0.77 \pm 0.19$, 
showing a similar pattern as $|a_1^{D_sK}|$, although with larger uncertainty. We hope that the experimental and theoretical precisions for this mode can be improved in the future. 
 
 In view of the intriguing results for $\gamma$ and the picture following from the branching ratios, we have generalised the analysis
 of the $B^0_s\to D_s^\mp K^\pm$ system to allow for NP contributions with new sources of CP violation. Employing a model-independent parametrisation,we have developed a formalism for the CP-violating observables to include also possible effects from direct CP violation, resulting finally in an effective angle $\gamma_{\rm eff}$ which differs from $\gamma$ through a CP-violating NP phase shift. We have also 
 generalised the expressions of the branching ratios of the underlying decay channels correspondingly. 
 
 In our NP analysis of the experimental data, we assume that CP-conserving phase differences vanish, which results in a setting consistent with the assumption made by the LHCb collaboration when measuring the observables. Applying our formalism, we have 
 calculated correlations between the NP parameters of the $b\to c \bar u s$ and $b \to u \bar c s$ quark-level transitions. 
 Interestingly, we obtain strongly correlated NP contributions with potentially large CP-violating phases. Moreover, we find 
 that NP contributions as small as about $30\%$ of the SM amplitudes could accommodate the current data. The construction of specific NP models faces challenges with high-$p_{\rm T}$ collider data for direct NP searches by the ATLAS and CMS collaborations.
 
 The strategy presented in this paper sets the stage for analyses of future measurements of the $B^0_s\to D_s^\mp K^\pm$ 
 system, employing also semileptonic
 $\bar{B}^0_s\to D_s^+ \ell^-\bar{\nu}_\ell$ and $\bar{B}^0_s\to K_s^+ \ell^-\bar{\nu}_\ell$ decays. It will be exciting to see how the
 data will evolve once we are moving to higher and higher precision. The burning question is whether we will finally be able
 to establish new sources of CP violation in the $B^0_s\to D_s^\mp K^\pm$ decays, which have been key players in this field 
 since the pioneering days of $B$ physics.

\vspace*{0.5truecm}


\section*{Acknowledgements}
We would like to thank Ruben Jaarsma, Philine van Vliet and Kristof De Bruyn for useful discussions. 
This research has been supported by the Netherlands Organisation for Scientific Research (NWO).


\newpage

\end{document}